\documentclass[fleqn,usenatbib]{mnras}

\usepackage{newtxtext,newtxmath}

\usepackage[T1]{fontenc}

\DeclareRobustCommand{\VAN}[3]{#2}
\let\VANthebibliography\thebibliography
\def\thebibliography{\DeclareRobustCommand{\VAN}[3]{##3}\VANthebibliography}

\usepackage{graphicx}	
\usepackage{amsmath}	
\usepackage{comment}
\usepackage[dvipsnames]{xcolor}
\usepackage{color}
\usepackage{bm}

\graphicspath{{./picture/}}

\newcommand{\be}{\begin{equation}}
\newcommand{\ee}{\end{equation}}
\newcommand{\bea}{\begin{eqnarray}}
\newcommand{\eea}{\end{eqnarray}}
\newcommand{\nn}{\nonumber}

\newcommand{\krb}[1]{\left(#1\right)}
\newcommand{\kmb}[1]{\left\{#1\right\}}

\newcommand{\kag}[1]{\langle#1\rangle}
\newcommand{\bi}{\begin{itemize}}
\newcommand{\ei}{\end{itemize}}

\newcommand{\fr}[2]{\frac{#1}{#2}}
\newcommand{\norm}[3]{\left(\frac{#1}{#2}\right)^{#3}}
\newcommand{\normi}[2]{\left(\frac{#1}{#2}\right)}

\newcommand{\Msun}{\ \mathrm{M}_{\odot}} 
 
\newcommand{\nH}{n_{\rm H}} 
\newcommand{\cc}{\ \mathrm{cm}^{-3}}
\newcommand{\cm}{\ \mathrm{cm}}

\newcommand{\kms}{\ \mathrm{km}~\mathrm{s}^{-1}}

\newcommand{\eV}{\ \mathrm{eV}}

\newcommand{\pc}{\ \mathrm{pc}}

\newcommand{\yr}{\ \mathrm{yr}}
\newcommand{\Myr}{\ \mathrm{Myr}}

\newcommand{\K}{\ {\rm K}}
\newcommand{\LEdd}{L_{\rm Edd}}
\newcommand{\tEdd}{t_{\rm Sal}}

\newcommand{\mdotEdd}{\dot{M}_{\rm Edd}}

\newcommand{\mdotBH}{\dot{M}_{\rm BH}}
\newcommand{\mdotBondi}{\dot{M}_{\rm Bondi}}

\newcommand{\mdotBHL}{\dot{M}_{\rm BHL}}
\newcommand{\mdotBHLII}{\dot{M}_{\rm BHL,II}}
\newcommand{\mBHdot}{\dot{M}_{\rm BH}}

\newcommand{\kB}{k_{\rm B}}

\newcommand{\mH}{m_{\rm H}}

\newcommand{\Mgas}{M_{\rm gas}}

\newcommand{\Msuny}{\ \mathrm{M}_\odot \mathrm{yr}^{-1}}

\newcommand{\erg}{\ \mathrm{erg}}
\newcommand{\x}{\times }
\newcommand{\Mhalo}{M_{\rm halo}}
\newcommand{\Tvir}{T_{\rm vir}}
\newcommand{\rvir}{r_{\rm vir}}
\newcommand{\Rvir}{R_{\rm vir}}
\newcommand{\vvir}{v_{\rm vir}}

\newcommand{\rdisc}{r_{\rm disc}}

\newcommand{\mBH}{M_{\rm BH}}
\newcommand{\mBHzero}{M_{\rm BH}(0)}

\newcommand{\mBHsan}{M_{\rm BH,3}}
\newcommand{\Mcl}{M_{\rm cl}}
\newcommand{\Rcl}{R_{\rm cl}}

\newcommand{\cs}{c_{\rm s}}

\newcommand{\rBondi}{r_{\rm Bondi}}

\newcommand{\rBHL}{r_{\rm BHL}}

\newcommand{\lJ}{\lambda_{\rm J}}

\newcommand{\tff}{t_{\rm ff}}
\newcommand{\tlife}{t_{\rm life}}
\newcommand{\tcool}{t_{\rm cool}}
\newcommand{\aJ}{a_{\rm J}}
\newcommand{\aff}{a_{\rm ff}}
\newcommand{\aBondi}{a_{\rm Bondi}}
\newcommand{\sigmaT}{\sigma_{\rm T}}
\newcommand{\rStr}{r_{\rm Str}}
\newcommand{\aB}{\alpha_{\rm B}}
\newcommand{\hnu}{h\nu}
\newcommand{\s}{~{\rm s}}
\newcommand{\nHII}{n_{\rm H, II}}
\newcommand{\TII}{T_{\rm II}}
\newcommand{\lsim}{\lesssim}
\newcommand{\gsim}{\gtrsim}
\newcommand{\MJ}{M_{\rm J}}

\newcommand{\Rar}{\Rightarrow}
\newcommand{\LRar}{\Leftrightarrow}
\newcommand{\Mtot}{M_{\rm tot}}
\newcommand{\vR}{v_{\rm R}}
\newcommand{\rhoII}{\rho_{\rm II}}
\newcommand{\vII}{v_{\rm II}}
\newcommand{\csII}{c_{\rm s,II}}
\newcommand{\vin}{v_{\rm in}}
\newcommand{\vesc}{v_{\rm esc}}
\newcommand{\nclump}{n_{\rm cl}}

\newcommand{\bmax}{b_{\rm max}}

\newcommand{\tcross}{t_{\rm cross}}

\newcommand{\fshell}{f_{\rm shell}}
\newcommand{\fDF}{f_{\rm DF}}

\newcommand{\rmd}{{\rm d}}
\newcommand{\nHmin}{n_{\rm H, min}}
\newcommand{\nHmax}{n_{\rm H, max}}

\newcommand{\sigmatr}{\sigma_{\rm trap}}
\newcommand{\fcl}{f_{\rm cl} }
\newcommand{\rperi}{r_{\rm peri}}
\newcommand{\tp}{t_{\rm p}}
\newcommand{\Pcl}{P_{\rm cl}}
\newcommand{\nHbar}{\bar{n}_{\rm H}}
\newcommand{\mBHfin}{M_{\rm BH,fin}}
\newcommand{\ft}{\text{--}}
\newcommand{\neII}{n_{\rm e,II}}
\newcommand{\muII}{\mu_{\rm II}}
\newcommand{\cmcs}{~\cm^3 \s^{-1}}
\newcommand{\Lv}{L_\nu}

\newcommand{\Dtdivsuper}{\Delta t_{\rm div}}
\newcommand{\Dtdoublesuper}{\Delta t_{\rm double}}
\newcommand{\tSC}{t_{\rm SC}}
\newcommand{\bmx}{{\bm x}}

\newcommand{\bmv}{{\bm v}}
\newcommand{\facc}{f_{\rm acc}}
\newcommand{\mure}{\mu_{\rm re}}
\newcommand{\Mach}{\mathcal{M}}
\newcommand{\Machturb}{\mathcal{M}_{\rm turb}}
\newcommand{\bturb}{b_{\rm turb}}

\newcommand{\rhoshell}{\rho_{\rm shell}}
\newcommand{\Drshell}{\Delta r_{\rm shell}}

\newcommand{\exponentHS}{\exp{\left\{-\fr{\left(\ln\left(\frac{n_{\rm H}}{\bar{n}_{\rm H}}\right) + \fr{\sigma^2}{2}\right)^2}{2\sigma^2}\right\} }}
\newcommand{\nII}{n_{\rm II}}
\newcommand{\Mdotcl}{\dot{M}_{\rm cl}}
\newcommand{\cshell}{c_{\rm shell}}
\newcommand{\vturb}{v_{\rm turb}}
\newcommand{\rhomin}{\rho_{\rm min}}

\defcitealias{Kiyuna+2023}{K23}

\title[Super-Eddington Growth Ceiling]{Super-Eddington Growth Ceiling: Analytic Constraints on the Rapid Growth of Light-Seed Black Holes in Massive Clumps}

\author[M. Kiyuna]{
Masaki Kiyuna$^{1}$\thanks{E-mail: kiyuna@astr.tohoku.ac.jp}
\\
$^{1}$Astronomical Institute, Graduate School of Science, Tohoku University, Aoba, Sendai 980-8578, Japan
}

\date{Accepted XXX. Received YYY; in original form ZZZ}

\pubyear{2025}

\begin{document}
\label{firstpage}
\pagerange{\pageref{firstpage}--\pageref{lastpage}}
\maketitle

\begin{abstract}
The existence of $\sim10^{7\ft8}\Msun$ supermassive black holes at $z\gsim8$ challenges conventional growth channels.  One attractive possibility is that light seeds ($\mBH\lsim10^{3}\Msun$) undergo short, super-Eddington episodes when they cross, and are captured by, dense massive gas clumps.  We revisit this ``BH-clump-capture'' model using analytic arguments supported by toy-model simulations that follow Bondi-scale inflow, radiative feedback, gas dynamical friction and the recently discovered forward acceleration effect caused by the ionised bubble.
For substantial growth the black hole must remain trapped for many dynamical times, which imposes three simultaneous constraints.  The clump must be heavier than the black hole (mass doubling condition); its cooling time must exceed the super-Eddington growth time (lifetime condition); and dynamical friction must dominate shell acceleration (BH-trapping condition).  These requirements confine viable clumps to a narrow density-temperature region, $n_{\rm H}\simeq10^{7\ft8}\cc$ and $T\simeq(2\ft6)\times10^{3}\K$, for a $10^{3}\Msun$ seed.
Even inside this sweet spot a $10^{3}\Msun$ seed grow up at most $4\times10^{3}\Msun$; the maximum growth ratio $\mBHfin/\mBH$ falls approximately as $\mBH^{-0.4}$ and is negligible once $\mBH\gsim10^{4}\Msun$.  The forward-acceleration effect is essential, expelling the black hole whenever photon trapping fails.  We conclude that BH-clump-capture model, and potentially broad super-Eddington models, cannot produce the $>10^{4}\Msun$ seeds required for subsequent Eddington-limited growth, suggesting that alternative pathways, such as heavy seed formation, remain necessary.
\end{abstract}



\section{INTRODUCTION}\label{sec:introduction}

Supermassive black holes (SMBHs), with masses ranging from $10^6$ to $10^{10}\Msun$, are found at the centres of most galaxies, yet their origin remains one of the major unsolved problems in astrophysics \citep{Inayoshi+2020, Regan&Volonteri2024}. The discovery of quasars powered by $\sim10^{9\ft10}\Msun$ SMBHs at redshifts $z > 6$ \citep{Fan+2006, Mortlock+2011, Banados+2018, Wang+2021} suggests that such massive objects had already formed within the first $\sim1$ billion years after the Big Bang. More recently, observations with the James Webb Space Telescope (JWST) have revealed the presence of SMBHs with masses of $10^{6\ft8}\Msun$ at even earlier epochs, $z = 8$-10 \citep{Kovacs+2024, Maiolino+2024b, Larson+2024, Bogdan+2024, Matthee+2024, Greene+2024}, further deepening the mystery of how they assembled so quickly.

The growth of black holes (BHs) through gas accretion is generally constrained by the Eddington limit. To evolve from so-called ``light seed BHs'' with masses $\lesssim10^3\Msun$, such as remnants of Population~III (Pop III) stars, into the SMBHs observed at those high redshifts, a mass increase of about six orders of magnitude is required. However, even under continuous accretion at the Eddington rate, such dramatic growth cannot be achieved within the age of the Universe.

Moreover, high-resolution simulations demonstrate that BH seeds with masses $\lesssim10^5\Msun$ rarely sustain Eddington-limited accretion over the many orders of magnitude of growth required, due to radiative feedback from the accreting BH itself \citep{Milosavljevic2009, Milosavljevic+2009b, Park&Ricotti2011, Park&Ricotti2012, Jeon+2012, Smith+2018}. 
One reason is that, when the BH mass is still small, the Bondi radius is correspondingly small,
\be
\rBondi \sim 0.06\pc \normi{\mBH}{10^3\Msun} \norm{T}{10^4\K}{-1},
\ee
meaning that the BH's gravitational sphere of influence encloses only a limited amount of gas. As a result, the BH alone cannot efficiently gather gas from its surroundings.

To overcome this limitation, some scenarios assume that the BH resides at the center of a larger gravitational structure, such as a galaxy or stellar cluster, and accretes gas that is drawn in by the overall gravitational potential of the system. However, for this to occur, the BH must be delivered to and retained at the center of the host system. While sufficiently massive BHs can achieve this via dynamical friction, for BHs with $\mBH \lesssim 10^5\Msun$, the friction is generally too weak to bring them into the dense nuclear region \citep{Pfister+2019}. Even BHs as massive as $\sim10^8\Msun$ may continue to wander without settling \citep{Ma+2021}.
Reaching a mass of $\sim10^5\Msun$ is therefore a crucial threshold for enabling efficient, sustained growth \citep{Regan&Volonteri2024}.

The seeding and growth of SMBHs are believed to proceed through one of two promising pathways.
In the ``heavy-seed followed by Eddington-limited growth'' scenario, black holes with initial masses of $10^{5\ft6}\Msun$ are formed either through (1) direct collapse in atomic-cooling haloes where H$_2$ cooling is suppressed \citep{Omukai2001, Bromm&Loeb2003, Dijkstra+2008, Shang2010, Inayoshi&Omukai2012, HosokawaYork2013, Inayoshi&Omukai&Tasker2014, Sugimura+2014, Hirano2015, Chon+2016, Wise+2019, Latif+2022}, or (2) runaway collisions among stars and stellar-mass black holes in dense star clusters \citep{Portegies+2002, Omukai2008, Devecchi+2009, Katz+2015, Sakurai+2017, Sakurai+2019, Reinoso+2018, Reinoso+2020, Vergara+2021, Vergara+2023, Vergara+2024, Rantala+2024}.
These formation channels require rare environments that are both metal-poor and exposed to intense radiation fields, and thus may not be common enough to account for the observed abundance of SMBHs.
Note, however, that alternative heavy-seed formation processes which could operate under more typical conditions are also being actively explored \citep{Chon+2020,Chon+2025,Chiaki+2023, Kiyuna+2023, Kiyuna+2024}.

The alternative is the ``light-seed with intermittent super-Eddington growth'' scenario, in which stellar-mass seeds with its mass of $10^{1\ft3}\Msun$ grow rapidly by accreting gas at rates exceeding the Eddington limit \citep{Haiman&Loeb2001, Volonteri+2003, Yoo&Miralda-Escude2004, Volonteri&Rees2005}.
Radiation-hydrodynamic simulations of gas inflow near BHs have shown that such super-Eddington accretion can be sustained due to photon trapping and anisotropic outflows \citep{Begelman1979, Abramowicz1988, Ohsuga+2005, Sadowski+2009, PacucciVolonteriFerrara2015, Sadowski&Narayan2016, Sadowski+2016, Jiang+2019}.
In addition, \citet{Inayoshi+2016} identified an even more extreme ``hyper-Eddington'' regime, which occurs when the ionised bubble surrounding the BH remains confined within the Bondi radius.
In this case, the accretion flow is not impeded by radiative feedback from the ionised region, and accretion proceeds unimpeded.
The condition for this regime to occur is:
\be
\nH \gsim 10^6\cc~\norm{\mBH}{10^3\Msun}{-1} \norm{T}{10^4\K}{1/2}.
\ee
Under these conditions, a $10^3\Msun$ seed can grow to $\sim10^5\Msun$ within $\sim10$ Myr \citep[see also][]{Toyouchi+2020}.

While these simulations focus on the physics near or within the Bondi radius, several challenges remain outside this region, making super-Eddington accretion far from straightforward. Most simulations optimistically place the BH at the halo centre and assume smooth, spherically symmetric inflow \citep{Volonteri&Rees2005, Alexander&Natarajan2014, Volonteri+2015, Inayoshi+2016, Ryu+2016, Shi+2022}, thereby neglecting the clumpy and turbulent nature of realistic galactic environments.

In reality, however, enabling sustained super-Eddington accretion requires overcoming several fine-tuning problems related to the large-scale environment: 
\begin{itemize}
\item[(1)] Efficient accretion requires the BH to be located in a dense environment, such as the halo center, and remain there stably on Bondi scales. While dynamical friction is often considered the primary mechanism for this, it is typically too weak for light seeds to both bring them to the center and keep them there \citep{Pfister+2019, Ma+2021}.
\item[(2)] Although the BH gravity is negligible beyond its Bondi radius, the collapse of gas via its own self-gravity must still converge on the BH's location with Bondi-scale precision—an unlikely coincidence.
\item[(3)] The inflow must resist fragmentation despite the presence of significant angular momentum and turbulent motions. In reality, inflowing gas is typically clumpy, turbulent, and rotationally supported—preventing coherent, large-scale collapse onto the BH.
\end{itemize}
These conditions are rarely met simultaneously in realistic environments. 

One proposed way to alleviate some of these difficulties is to embed the BH in a nuclear disc, bulge, or dense stellar cluster, which can help maintain its central position and enhance gas inflow from larger radii \citep[e.g.][]{Alexander&Natarajan2014, Lupi+2014, Inayoshi+2022, Lupi+2024}. However, fragmentation within such systems remains a major obstacle. Accepting that fragmentation is inevitable, \citet{Lupi+2016} proposed an alternative pathway—{\it the BH-clump-capture model}—in which a wandering BH grows through episodic encounters with massive gas clumps.
Their simulations demonstrate that a $20\Msun$ BH embedded in a circum-nuclear disc (CND) can grow to $10^4\Msun$ through a single encounter with a massive gas clump. In this scenario, the BH is not fixed at the centre but dynamically interacts with orbiting clumps, introducing stochasticity to its growth. The results highlight the possibility that BHs can still grow rapidly via clump capture, even in environments where gas fragmentation is prevalent. This BH-clump-capture mechanism has since been investigated by several follow-up simulations across a range of physical scales \citep{Shi+2022, Shi+2024b, Massonneau+2023, Sassano+2023, Gordon+2024}.

When the mass of a clump exceeds that of the BH, the clump's physical properties, rather than the BH gravity, govern the conditions at the Bondi radius. The balance between pressure and gravity determines the clump's fragmentation scale and sets the gas density, temperature, and relative velocity that the BH encounters. Such configurations commonly arise in environments such as fragmented discs, dense star clusters, galaxy mergers, and turbulent flows, where the gas reservoir outweighs the seed BH. Understanding how clump properties regulate BH accretion is therefore crucial for assessing the viability of super-Eddington growth.

The BH-clump-capture model, however, still lacks a clear physical understanding. The pioneering simulations by \citet{Lupi+2016} adopted an accretion radius comparable to the clump size, far larger than the Bondi radius, which likely led to an overestimation of the accretion rate. Resolving the Bondi scale more accurately may reveal that a BH passing through a clump accretes only a small fraction of its gas. Moreover, their study explored only a limited set of parameters, leaving it unclear how massive a BH can ultimately grow via this mechanism. A more systematic evaluation is therefore needed.

Here, we examine whether the BH-clump-capture mechanism can grow a $\sim10^{3}\Msun$ seed to $\gtrsim10^{5}\Msun$.
We combine analytic arguments with toy-model simulations that incorporate the BH-clump orbital dynamics, Bondi-scale accretion, radiative feedback based on the framework of \citet{Park&Ricotti2013, Sugimura&Ricotti+2020}, gas dynamical friction, and forward acceleration by the ionised bubble shell \citep{Toyouchi+2020, Ogata+2024}.
This forward acceleration refers to the gravitational pull exerted by the dense shell at the outer edge of the ionised bubble surrounding the BH.  
As this upstream bubble sweeps up and accumulates the oncoming gas, its shell drags the BH further upstream.
We find that this shell-induced force dominates over dynamical friction, severely restricting the range of clump masses capable of trapping the BH. As a result, while clump capture can boost the BH mass up to $\mBH \lesssim 10^4\Msun$, it is unlikely to surpass the $\sim10^5\Msun$ threshold. The super-Eddington growth scenario for light seeds therefore appears insufficient to produce the massive BH seeds required to explain the most distant quasars.

This paper is organised as follows. Section~\ref{sec:concept} outlines the basic framework of the BH-clump-capture model and introduces the three key conditions required for sustained growth. Sections~\ref{sec:doubling_condition} and \ref{sec:lifetime_condition} analytically derive the ``mass doubling condition'' and ``cloud lifetime condition,'' respectively. Section~\ref{sec:trap_condition} presents the results of our toy-model simulations and derives the ``BH-trapping condition'' based on those results. Section~\ref{sec:discussion} discusses the implications of our findings and explores the potential effects of physical processes not included in our model. Finally, Section~\ref{sec:conclusion} summarises our conclusions.

\begin{figure*}
	\centering
	\includegraphics[bb=0 0 3200 1700,width=15cm,scale=0.2]{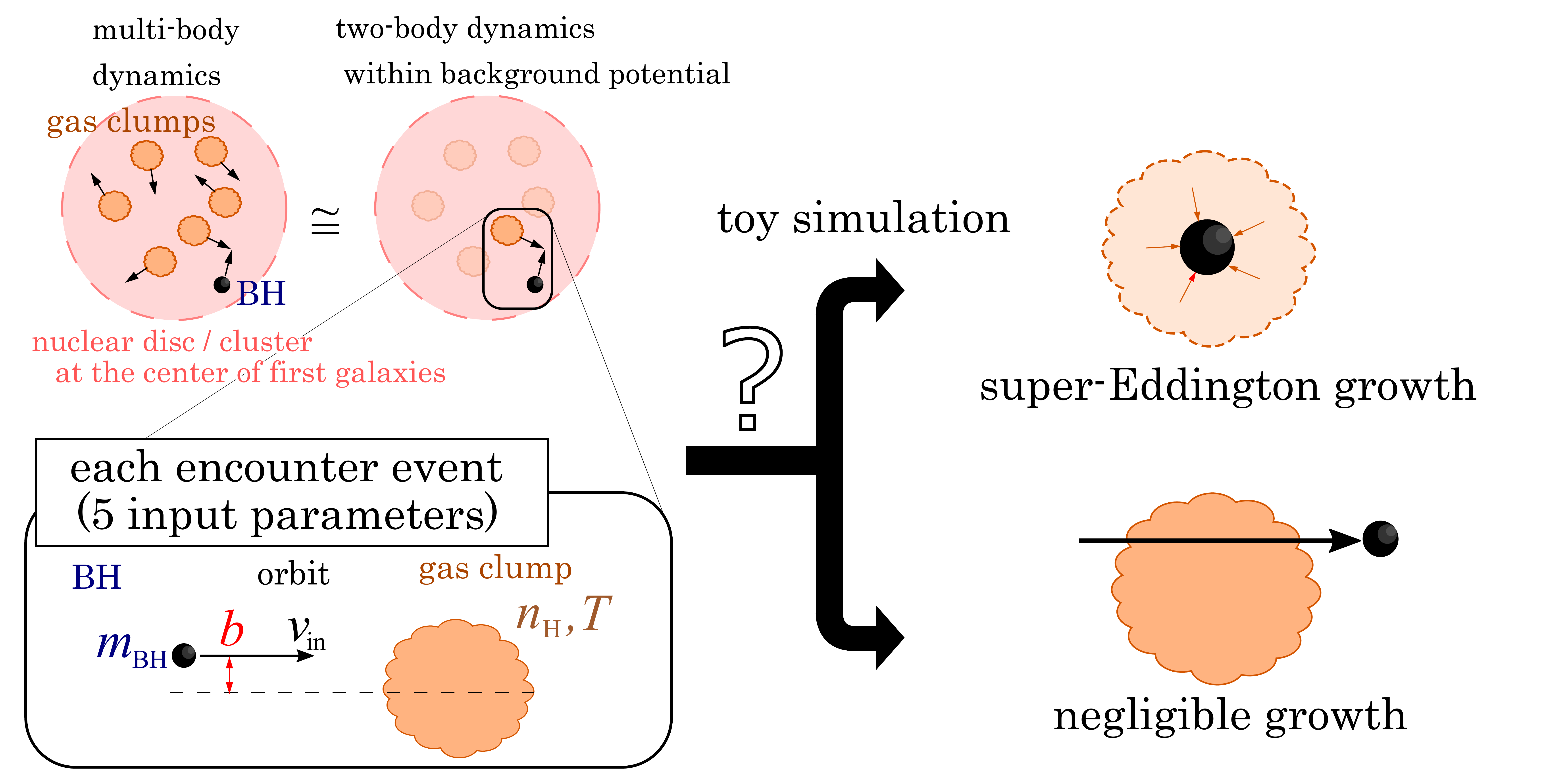}
\caption{A schematic illustration of the set-up and conceptual framework of our toy-model simulation, designed to examine the BH-clump interaction in the context of early galactic nuclei.
In the nuclear disc or cluster at the center of the first galaxies, numerous dense gas clumps are expected to form due to some effects, such as gravitational instabilities.
We model the interaction between a single gas clump and a BH as a two-body dynamical problem embedded within a static background gravitational potential.
This simplified approach allows us to determine whether the BH is captured by the clump and undergoes rapid, super-Eddington accretion, or escapes with negligible growth.
The toy model is specified by five input parameters: the initial BH mass $\mBH$, the gas number density $\nH$, the gas temperature $T$, the initial relative velocity $\vin$, and the impact parameter $b$.
By scanning these parameters, we delineate the conditions under which BH trapping and efficient growth occur.
}
    \label{fig:concept}
\end{figure*}

\section{Basic Framework and Three Conditions for BH-Clump Growth
}\label{sec:concept}

\subsection{Formulation of the BH-clump-capture model}
In this section we outline an analytical framework for the BH-clump-capture model (see Figure~\ref{fig:concept}).  
While we adopt several idealised assumptions for simplicity, we demonstrate that even under such favourable conditions, the model faces significant challenges in super-Eddington accretion.

Following \citet{Lupi+2016}, we consider a high-density CND at the centre of a halo and place a light seed BH within the disc. 
We assume that processes such as thermal instability, angular momentum transport, and turbulence fragment the gas into clumps.
Throughout this study, we adopt a fiducial seed mass of $\mBH = 10^3\Msun$, representative of a Pop III star remnant, either carried into the CND from minihaloes or formed \emph{in situ}.

When the BH encounters a clump and passes through its interior, dynamical friction and gas accretion couple the BH's orbital evolution to its mass growth. We approximate the interaction as a two-body problem, neglecting external perturbations. This approximation is adequate during the closest phase of the encounter. In the most extreme case, the BH could accrete the entire gas mass of the clump, increasing its mass to $\mBH + \Mcl$.

Each clump is modeled as a uniform sphere, and tidal deformation during the encounter is neglected. Assuming hydrostatic equilibrium, where thermal pressure balances self-gravity, the clump is characterised by its gas density $\nH$ and temperature $T$, which we treat as control parameters. The clump radius and mass are set to half the Jeans length and the Jeans mass, respectively:
\bea
\Rcl &=& \frac{\lJ(\nH, T)}{2} \equiv \frac{\sqrt{\pi}}{2}\frac{\cs}{\sqrt{G\rho}} =\norm{4\pi}{3}{-1/3}\aJ^{1/3}\frac{\cs}{\sqrt{G\rho}}\nn \\
     &\simeq& 2.0 \x  10^{-1} \pc~\norm{\nH}{10^7\cc}{-1/2}\norm{T}{10^4\K}{1/2}, \label{eq:Jeans_Length} \\
\Mcl &=& \MJ(\nH, T) \equiv \frac{4\pi}{3}\rho \Rcl^3 =  \aJ \frac{\cs^3}{\sqrt{G^3 \rho}} \nn \\
     &\simeq& 1.0 \x  10^4 \Msun~\norm{\nH}{10^7\cc}{-1/2}\norm{T}{10^4\K}{3/2}, \label{eq:Jeans_Mass}
\eea
where $\cs=\sqrt{\kB T/\mu\mH}$ and $\aJ=(4\pi/3)(\sqrt{\pi}/2)^3\simeq2.9$.  
We adopt a mean molecular weight of $\mu = 1.22$, and denote the clump mass explicitly as $\Mcl = \MJ(\nH, T)$. The statistical abundance of such clumps is discussed in Section~\ref{subsec:Statistics}.

The two-body orbits, confined to a two-dimensional plane, are fully characterised by the initial relative velocity $\vin$ and the impact parameter $b$. The system is therefore described by five parameters: $(\mBH, \nH, T)$, which define the BH-clump configuration, and $(\vin, b)$, which specify the encounter geometry. In this work, we delineate the regions of this parameter space that permit super-Eddington growth.

We assume metal-free gas and neglect both metal-line and H$_2$ cooling. The potential effects of these additional cooling channels are discussed in Section~\ref{sec:lifetime}.

\subsection{Three conditions for Super-Eddington BH Growth}\label{subsec:3_conditions}

Using the framework outlined above, we now identify the conditions under which a BH can undergo substantial super-Eddington growth. Here, ``substantial'' refers to an accretion episode that increases the BH mass by at least a factor of two. As demonstrated analytically in Section~\ref{sec:lifetime_condition}, a single passage through a clump results in only negligible mass accretion.
Therefore, substantial growth requires the BH to remain gravitationally bound within the clump for a period much longer than the crossing time, i.e., $\Delta t \gg \tcross$. This requirement imposes three additional constraints on the BH-clump system:

\vspace{0.5em}
\noindent
\textbf{i) Mass doubling condition:}  
The clump mass must exceed the BH mass; otherwise, the BH cannot double its mass in a single encounter. As discussed in Section~\ref{sec:doubling_condition}, this condition favours lower $\nH$ and higher $T$.

\vspace{0.5em}
\noindent
\textbf{ii) Clump lifetime condition:}  
The clump must survive longer than the BH accretion timescale.
We assume that the clump lifetime is comparable to its cooling time, i.e.\ $\tlife \sim \tcool$.
If this condition is not satisfied, the clump collapses or fragments before the BH can accrete significant mass, thereby halting its growth.
In particular, efficient Ly$\alpha$ cooling at $T \gtrsim 10^4\K$ places an upper limit of $T \lesssim 10^4\K$ for sustained growth.
A quantitative formulation of this criterion is derived in Section~\ref{sec:lifetime_condition}.

\vspace{0.5em}
\noindent
\textbf{iii) BH-trapping condition:}  
For a BH to be trapped within a clump and remain there throughout the accretion phase, the dynamical friction acting on the BH must overcome the forward acceleration exerted by the ionised bubble shell.
As shown in Section~\ref{sec:trap_condition}, this requirement is more stringent than the hyper-Eddington condition and favours high $\nH$ and low $T$.

We examine these three conditions sequentially in Sections~\ref{sec:doubling_condition}-\ref{sec:trap_condition}.
Sections~\ref{sec:doubling_condition} and~\ref{sec:lifetime_condition} derive the mass doubling and clump lifetime conditions, respectively, based on analytical arguments.
Section~\ref{sec:trap_condition} then introduces the BH-trapping condition, established via toy-model simulations.
These three, partially conflicting, requirements are simultaneously fulfilled only within a narrow sweet spot in the $\nH$-$T$ plane.

\section{MASS DOUBLING CONDITION}\label{sec:doubling_condition}
To ensure that the BH can at least double its mass during an encounter, we impose the mass doubling condition on the BH-clump system:
\bea
\Mcl(\nH, T) &\geq& \mBH \nn,  
\eea
which translates into the following upper limit on the gas density:
\bea
\nH &\leq& \fr{\aJ^{2} \kB^{3} T^{3}}{(\mu\mH)^{4} G^{3} \mBH^{2}} \nn \\
         &\simeq& 1.0 \x  10^{9}\cc~\norm{T}{10^{4}\K}{3}\,\mBHsan^{-2}, \label{eq:doubling_condition}
\eea
where $\mBHsan\equiv\mBH/10^{3}\Msun$.  
In Section~\ref{sec:trap_condition}, our toy-model simulation explores the region of parameter space that satisfies Equation~(\ref{eq:doubling_condition}).

When this condition is violated ($\Mcl<\mBH$),
the clump is subject to strong tidal forces from the BH, leading to significant distortion. In such cases, the assumption of a uniform, spherical clump structure breaks down.
Moreover, in this regime, the clump radius  
\bea
\fr{G\Mcl}{\Rcl} &\sim&\norm{4\pi}{3}{1/3}\aJ^{2/3}\cs^2\nn\\
\Rcl &\sim& \norm{4\pi}{3}{-1/3}\aJ^{-2/3}\normi{G\Mcl}{\cs^2}\simeq 0.3\normi{G\Mcl}{\cs^2}
\eea
becomes smaller than the BH Bondi radius, $\rBondi = G\mBH/\cs^{2}$.
Here we have used Equations~(\ref{eq:Jeans_Length}) and ~(\ref{eq:Jeans_Mass}).  
Although such a clump entering within $\rBondi$ could, in principle, be accreted at a transient super-Eddington rate, the corresponding mass gain falls short of the initial BH mass.
Furthermore, since the associated radiation can escape beyond $\rBondi$ without photon trapping, it is expected to suppress further inflow via standard Eddington-limited feedback.
We therefore regard accretion from clumps with $\Mcl < \mBH$ as incompatible with the steady photon-trapping regime described by \citet{Inayoshi+2016}, and we do not consider this regime further in the present work.

\section{CLUMP LIFETIME CONDITION}\label{sec:lifetime_condition}
In this section, we derive the condition that the clump must survive long enough to allow substantial BH growth.
Section~\ref{subsec:re-estimate_t_growth} estimates the characteristic timescale for BH mass increase via super-Eddington accretion.
Section~\ref{sec:lifetime} then evaluates the clump lifetime, comparing it to the required growth time to obtain the clump lifetime condition.

\subsection{BH growth timescale}\label{subsec:re-estimate_t_growth}

We now estimate the BH growth timescale under the assumption that the photon-trapping condition is satisfied, allowing sustained super-Eddington accretion.
Our accretion model is based on recent radiation-hydrodynamic simulations \citep{Hu+2022, Toyouchi+2024}, and is described in detail in Appendix~\ref{Appx:BH_accretion_model}.

In a clump where the photon-trapping condition is satisfied, the mass accretion rate onto the BH can be estimated as follows.
\citet{Hu+2022} and \citet{Toyouchi+2024} argue that, when gas accretes through a disc at a rate far exceeding the Eddington limit, forming a so-called slim disc structure, most of the inflowing material is expelled as outflows, and only a small fraction reaches the BH.
The effective BH accretion rate can be approximated by
\bea
\mdotBH \simeq \left(\mdotBHL\,\mdotEdd\right)^{1/2}
             \lsim\left(\mdotBondi\,\mdotEdd\right)^{1/2},          \label{eq:mdot_eff_main}
\eea
where $\mdotBondi$ and $\mdotBHL$ are the Bondi and Bondi-Hoyle-Littleton accretion rates, respectively:
\bea
\mdotBondi &=& \frac{\pi G^{2}\mBH^{2}\rho}{\cs^{3}}
             = \aBondi\,\frac{\mBH}{\Mcl}\frac{\mBH}{\tff}, \label{eq:Bondi_acc_rate}                     \\
\mdotBHL   &=& \frac{4\pi G^{2}\mBH^{2}\rho}{(\cs^{2}+v^{2})^{3/2}}
             = \frac{4\aBondi}{\left[1+(v/\cs)^{2}\right]^{3/2}}
               \frac{\mBH}{\Mcl}\frac{\mBH}{\tff}.                               \label{eq:BHL_main}
\eea
The free-fall time, set by the self-gravity of the clump, is given by
\bea
\tff\equiv\sqrt{\frac{3\pi}{32G\rho}} = \fr{\aff}{\sqrt{G\rho}},                                               \label{eq:tff_main}
\eea
where $\aff=0.54$ and $\aBondi=\pi\aJ\aff\simeq4.9$ are dimensionless coefficients.
Equations~(\ref{eq:Bondi_acc_rate}),~(\ref{eq:BHL_main}) show that a larger clump mass leads to lower accretion rates.  Physically, Bondi accretion operates inside the Bondi radius, where BH gravity dominates gas pressure.  As the clump mass increases, stronger self-gravity requires higher pressure and temperature, which shrinks the Bondi radius and thus reduces the accretion rate.

The Eddington accretion rate is defined as
\bea
\mdotEdd &=& \frac{1-\eta}{\eta}\frac{4\pi G \mH}{c\sigmaT}\mBH
           \equiv\frac{\mBH}{\tEdd}                                                   \nn\\
         &=& 2.0\times10^{-5}\Msuny\,\mBHsan,                                         \\
\tEdd    &=& 5.0\times10\Myr,                                                         \label{eq:tEdd_main}
\eea
where we adopt a radiative efficiency $\eta = 0.1$, and $\sigmaT$ is the Thomson cross-section for free electrons.

Substituting Equations~(\ref{eq:mdot_eff_main})-(\ref{eq:tEdd_main}) and assuming constant $\rho$ and $T$, we obtain
\bea
\mdotBH &\lsim& \aBondi^{1/2}
            \left(\frac{\mBH}{\Mcl}\right)^{1/2}
            \frac{\mBH}{(\tEdd\tff)^{1/2}},                                          \label{eq:mdot_param_main}\\
\mBH(t) &\lsim& \frac{\mBHzero}
                 {\left[1-
                   \frac{\aBondi^{1/2}}{2}
                   \left(\frac{\mBHzero}{\Mcl}\right)^{1/2}
                   \frac{t}{(\tEdd\tff)^{1/2}}\right]^{2}}.                           \label{eq:m_evo_main}
\eea
The divergence time $\Dtdivsuper$, defined as the time when the BH mass formally diverges, $\mBH(\Dtdivsuper) \to \infty$, is given by
\bea
\Dtdivsuper &\gsim& 1.1
                     \left(\frac{\Mcl}{\mBH}\right)^{1/2}
                     (\tEdd\tff)^{1/2}                                                \nn\\
           &=& 3.1\Myr\,
               \norm{\nH}{10^{7}\cc}{-1/2}
               \norm{T}{10^{4}\K}{3/4}\,
               \mBHsan^{-1/2}.\nn\\                                                        &&\label{eq:Dt_div_num}
\eea

The photon-trapping condition $\rStr<\rBHL$ 
can be recast as a condition on timescales (see Equation~(\ref{eq:hyper-Eddington_condition_tff})):
\bea
\tEdd &>& 8.2\times10
        \left(\frac{\Mcl}{\mBH}\right)\tff\nn\\
        &&\x \sqrt{1+\frac{v^{2}}{\cs^{2}}}
        \norm{\TII}{4\times10^{4}\K}{0.4}
        \normi{\TII/4\times10^{4}\K}{T/10^{4}\K},                                    \label{eq:trap_timescale}
\eea
where $\TII$ is the temperature of the ionized bubble surrounding the BH.
Combining this with Equation~(\ref{eq:Dt_div_num}), we obtain a lower limit on the BH divergence time:
\bea
\Dtdivsuper &>& 10
               \left(\frac{\Mcl}{\mBH}\right)\tff\nn\\
               &&\x \norm{\TII}{4\times10^{4}\K}{0.2}
               \norm{\TII/4\times10^{4}\K}{T/10^{4}\K}{1/2}.
               \label{eq:BH_div}
\eea
The time required for the BH mass to double, 
$\mBH(\Dtdoublesuper)=2\mBH(0)$, is also given by
\bea
\Dtdoublesuper > 3
                \left(\frac{\Mcl}{\mBH}\right)\tff.\label{eq:Dt_double}
\eea
These expressions define the characteristic timescales necessary for significant BH mass growth in the super-Eddington regime.

The time required for the BH to pass through the clump is
\bea
\tcross \sim \sqrt{\frac{2\Rcl^{3}}{G\Mcl}}
          = \frac{4}{\pi}\tff
          \sim\tff.                                                                   \label{eq:tcross_main}
\eea
Because the mass doubling condition requires $\Mcl>\mBH$, Equations~(\ref{eq:Dt_double}) and (\ref{eq:tcross_main}) imply
\bea
\tcross<\Dtdoublesuper.\label{eq:negligible_mass_growth}
\eea
Hence, a BH crossing a clump experiences negligible mass growth.
As mentioned in Section~\ref{subsec:3_conditions}, substantial growth, such as mass doubling, requires the BH to be gravitationally captured and remain bound to the clump for multiple free-fall times.
This in turn imposes additional conditions on the clump's physical state and the nature of the BH-clump interaction, i.e., BH-trapping condition (Sec. 5).

\subsection{Clump lifetime}\label{sec:lifetime}
We begin by summarising the basic properties of self-gravitating gas clumps. We then derive the \emph{clump lifetime condition}, which requires that a clump survive longer than the BH growth timescale derived in Section~\ref{subsec:re-estimate_t_growth}.

Gas clumps are assumed to form via such processes as fragmentation induced by thermal instability, angular momentum transport, turbulence, or converging flows.
Some clumps may initially deviate from dynamical equilibrium, with masses either above or below the Jeans mass determined by their density $\nH$ and temperature $T$.
Such clumps typically relax on the shorter of the free-fall time $\tff$, or the sound-crossing time $\tSC \simeq \Rcl/\cs$.
Since BH growth during this brief relaxation phase is negligible when $\Mcl > \mBH$ (see Equation~(\ref{eq:BH_div})), we safely regard clumps in the BH-clump-capture model are in dynamical equilibrium.

Once dynamical equilibrium is established, a clump cannot remain indefinitely at arbitrary values of $(\nH, T)$.
Radiative cooling drives thermal instability, restricting the clump lifetime to the cooling time
\bea
\tcool = \frac{1.5\,\nH\kB T}{\nH^{2}\Lambda(T)},                                         \label{eq:tcool_def}
\eea
where $\Lambda(T)$ is the radiative cooling function.  

The subsequent evolution of the clump depends on the relative magnitudes of the free-fall time $\tff$ and the cooling time $\tcool$:
\begin{itemize}

\item[\rm{(i)}] If $\tcool < \tff$, the clump cools efficiently prior to collapse, reducing its temperature and hence the Jeans mass. This triggers further collapse and fragmentation, with the initial thermodynamic state preserved only for a short duration $\tlife \sim \tcool (< \tff)$.

\item[\rm{(ii)}] If $\tcool \simeq \tff$, the clump collapses on a dynamical timescale and promptly forms stars \citep{Larson1969}. In this case, the clump lifetime is $\tlife \sim \tff$.

\item[\rm{(iii)}] If $\tcool > \tff$, the clump evolves quasi-statically: radiative cooling removes thermal energy while maintaining dynamical equilibrium, gradually increasing the gas temperature.
In primordial gas, this temperature rise leads to a sharp decrease in $\tcool$ as $T$ approaches $\sim 10^{4}\K$, ultimately transitioning the system into regime~(ii).
The initial clump state is thus maintained for a duration $\tlife \sim \tcool (> \tff)$.
\end{itemize}
Given the above, we approximate the clump lifetime by the cooling time: $\tlife = \tcool$.

To enable substantial BH growth via the BH-clump-capture model, the clump must survive for at least the BH accretion timescale. This requirement defines the \emph{clump lifetime condition}:
\bea
\tcool>\Dtdivsuper,  \label{eq:lifetime_condition_easy}
\eea
In practice, this condition implies that the clump must remain in quasi-static equilibrium, i.e., in regime~(iii), where $\tcool > \tff$.

Substituting Equation~(\ref{eq:BH_div}) into the lifetime condition (\ref{eq:lifetime_condition_easy}) and explicitly expressing the dependence on the cooling function $\Lambda(T)$ yields
\bea
\Lambda(T)
<2.6\times10^{-33}\,\erg\,\cmcs\,\mBHsan
\norm{\TII}{4\times10^{4}\,\K}{-0.7}.                                                   \label{eq:lifetime_Lambda}
\eea
Adopting the primordial atomic cooling function, which is dominated by Ly$\alpha$ emission below $10^{4}\K$, this inequality translates into a constraint on the temperature:
\bea
T<6.5\times10^{3}\,\K\,
\mBHsan^{0.02}\,
\norm{\TII}{4\times10^{4}\,\K}{-0.01}.                                             \label{eq:lifetime_condition}
\eea

Throughout, we have implicitly assumed that metal-line and H$_2$ cooling are negligible.
If either cooling mechanism were efficient, the gas could cool to significantly lower temperatures, increasing $\Lambda(T)$ and tightening the threshold in Equation~(\ref{eq:lifetime_condition}) by requiring even lower values of $T$.
Such enhanced cooling would thus make the lifetime condition more stringent, further constraining the viability of the BH-clump-capture scenario.
For instance, considering H$_2$ cooling in the LTE limit \citep{Glover2015_II}, Equation~(\ref{eq:lifetime_Lambda}) requires the gas temperature to be $T \lesssim 7 \times 10^2\K$ for a gas with number density $\nH = 10^7\cm^{-3}$ and molecular hydrogen fraction $y_{\rm H_2} = 10^{-4}$.

\section{BH-TRAPPING CONDITION}\label{sec:trap_condition}
In this section, we derive the condition under which a BH becomes \emph{trapped} within a clump.
The central issue is the competition between two forces acting on the BH: (1) dynamical friction exerted by the clump gas and (2) the forward acceleration induced by the ionised shell surrounding the BH.
Trapping occurs when the former dominates over the latter, allowing the BH to be gravitationally captured by the clump.

Both forces depend sensitively on the instantaneous relative velocity between the BH and the clump, which itself evolves under the influence of these same forces.
As such, the outcome is inherently coupled to the details of BH accretion and radiative feedback processes (see Appendices~\ref{Appx:Park-Ricotti_model}, \ref{Appx:BH_accretion_model}, and \ref{Appx:Kinematics} for full descriptions).
Due to the non-linear and time-dependent nature of these interactions, the BH-trapping condition cannot be derived analytically in the same straightforward manner as the mass doubling or clump lifetime conditions.

We address the BH-trapping problem using a suite of toy-model simulations, systematically scanning the relevant parameter space to identify the regions where BH capture occurs.
Based on the simulation results, we then construct an analytic formulation for the BH-trapping condition.
  
Section~\ref{sec:methods} outlines the setup of the toy model.
Sections~\ref{subsec:fiducial}-\ref{subsubsec:BH_mass_dependency} present the simulation results and derive an analytic expression for the BH-trapping criterion.
We begin in Section~\ref{subsec:fiducial} with a fiducial case assuming a BH mass of $\mBH = 10^{3}\Msun$ and nearly radial initial orbits ($\vin \rightarrow 0$, $b \rightarrow 0$), exploring a wide range of $(\nH, T)$.
Section~\ref{subsubsec:v_b_dependency} then examines how the trapping condition changes with the initial orbital parameters $(\vin, b)$, while Section~\ref{subsubsec:BH_mass_dependency} investigates the dependence on BH mass $\mBH$.

\subsection{Toy model setup}\label{sec:methods}
In our toy model, the clump is represented as a uniform, self-gravitating gas sphere with radius $\Rcl$, while the BH is treated as a point mass. 
We integrate the coupled evolution of their orbital motion and the BH mass using a fifth-order Runge-Kutta-Fehlberg method with adaptive time stepping.

The master equations governing the relative position, velocity, BH mass, and clump mass are
\bea
\dot{\bmx} &=& \bmv,\\
\dot{\bmv} &=& -\frac{G\Mtot(<r)}{r^{3}}\bmx  \nn\\
&&- \theta(\Rcl - r)\,\mure^{-1}\bigl(\fDF + \facc - \fshell\bigr)\frac{\bmv}{v},\\
\mdotBH &=&
\begin{cases}
    (\mdotEdd\,\mdotBHL)^{1/2} & (\rStr < \rBHL),\\
    \mdotEdd & (\mdotBHLII > \mdotEdd),\\
    \mdotBHLII & \text{otherwise},
\end{cases}\label{eq:acc_rate_method}\\
\Mdotcl &=& -\mdotBH,
\eea
where $\bmx$ and $\bmv$ are the relative position and velocity. The term $\fDF$ represents the dynamical friction, $\facc$ is the back-reaction force from mass accretion, and $\fshell$ is the forward acceleration due to the ionised shell.  
The three BH accretion rates in Equation~(\ref{eq:acc_rate_method}) correspond to the super-Eddington, Eddington-limited, and sub-Eddington regimes, respectively, and are specified as functions of $(\nH, T, \mBH, v)$.
For further details, see Appendix~\ref{Appx:Kinematics} (kinematics) and Appendices~\ref{Appx:Park-Ricotti_model},~\ref{Appx:BH_accretion_model} (accretion prescription).
Total mass is conserved such that $\Mcl(t)+\mBH(t)=\Mcl(t_{0})+\mBH(t_{0})$, and we scale the gas density proportionally with clump mass, i.e., $\nH(t)\propto\Mcl(t)$, while holding the clump radius fixed.  
A more realistic treatment of gas density would require solving the energy equation to track gas density and pressure evolution, but this is beyond the scope of the present toy model. This simplification affects outcomes only when the BH is captured and does not alter whether it is captured.

In each run the moment at which the BH first reaches $r=\Rcl$ is set to $t=0$.  
The integration terminates when the time exceeds the clump lifetime ($t\ge\tcool$) or when a super-Eddington phase commences and subsequently ceases.  
The BH mass at that moment is recorded as the final value $\mBHfin$.

\subsection{Fiducial runs}\label{subsec:fiducial}

We first fix the BH mass at $\mBH = 10^3\Msun$ and adopt idealised initial conditions that favour trapping, namely $\vin \to 0$ and $b \to 0$, corresponding to nearly zero initial mechanical energy $E \to 0$.
The clump density is varied over the range $\nH = 10^{2\ft10}\cc$ and the temperature over $T = 10^{1\ft5}\K$ to explore the parameter space.
Both quantities are sampled logarithmically as $10^{n/16}$, where $n$ is an integer.
For computational convenience, we adopt $\vin = 0.001\sqrt{2G\Mtot/\Rcl}$ and $b = 0.01\Rcl$ in each simulation.
These values are sufficiently small to approximate the limit $\vin \to 0$ and $b \to 0$.

\subsubsection{Time evolution in representative cases}\label{subsubsec:representative_runs}

\begin{figure}
    \includegraphics[bb=0 0 670 630,width=10cm,scale=0.2]{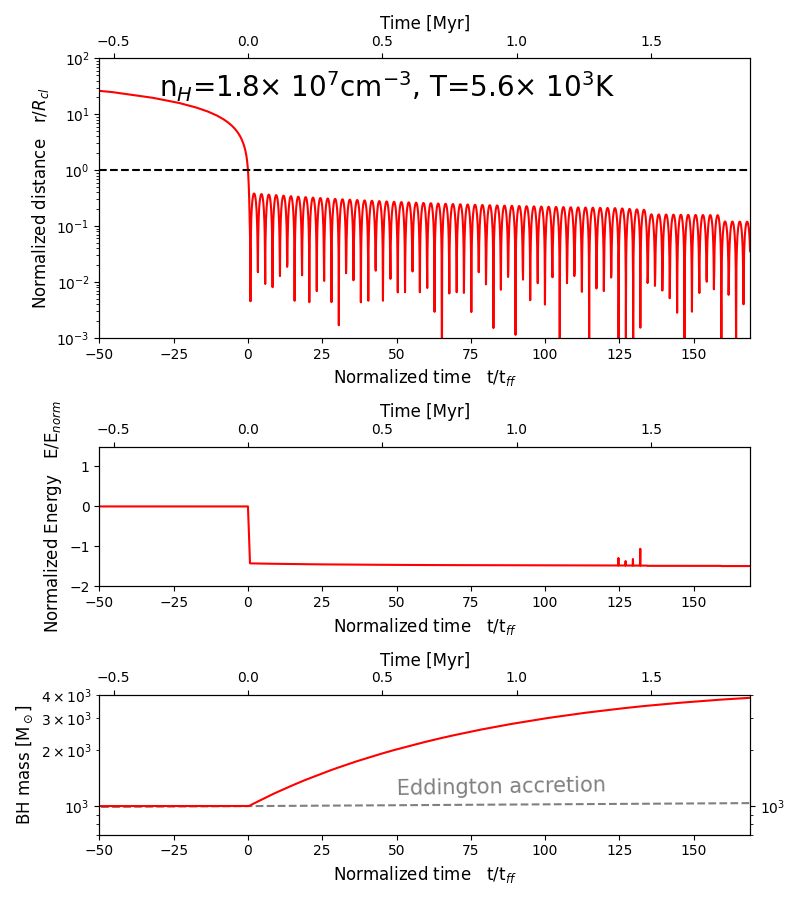}
\caption{
Time evolution of key physical quantities in a representative run in which the BH is successfully trapped within the clump, selected from the fiducial setup ($\mBH = 10^3\Msun,\ \vin \to 0,\ b \to 0$).  
The clump parameters are $\nH = 1.8 \times 10^7\cc$ and $T = 5.6 \times 10^3\K$.  
The horizontal axis shows time normalised by the free-fall timescale $\tff$, with $t=0$ defined as the moment when the BH first enters the clump ($r=\Rcl$). The evolution is shown up to the end of the super-Eddington accretion phase.  
\textbf{Top:} Radial distance of the BH from the clump centre, normalised by $\Rcl$ (red solid line); the black dashed line marks $r = \Rcl$.  
\textbf{Middle:} Mechanical energy of the BH, normalised by $E_{\rm norm} \equiv \sqrt{G(\Mcl+\mBH)/\Rcl}$.  Negative values indicate the BH is gravitationally bound to the clump.  
\textbf{Bottom:} BH mass. The red solid line shows the actural growth in the simulation, while the grey dashed line represents Eddington-limited growth for comparison.
}
\label{fig:tevo_o}
\end{figure}
\begin{figure}
    \includegraphics[bb=0 0 670 630,width=10cm,scale=0.2]{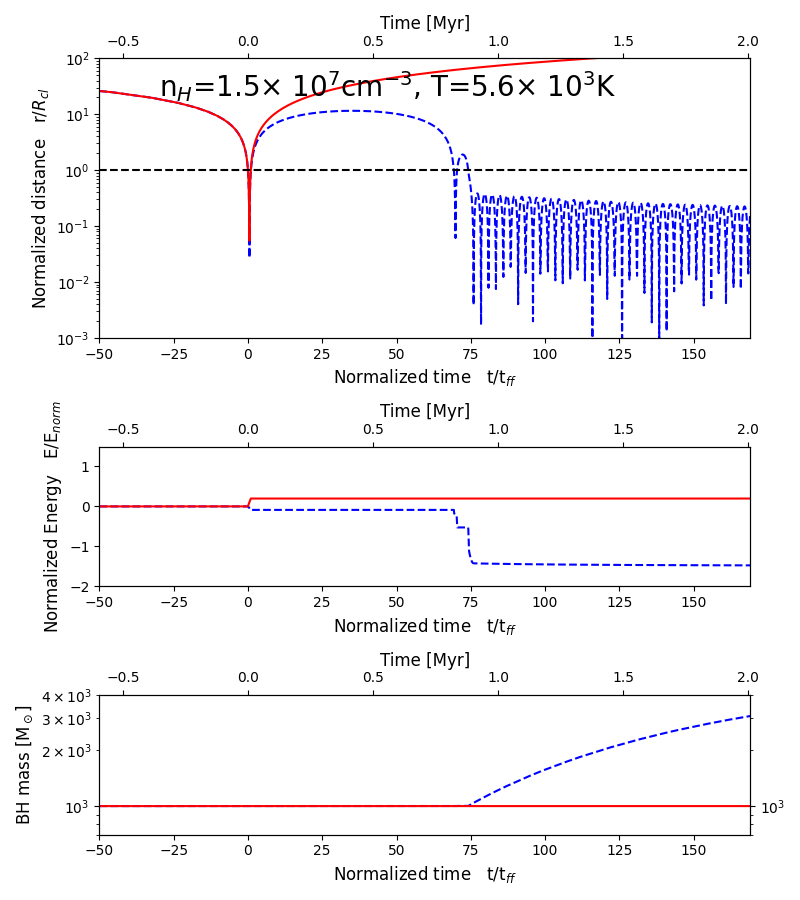}
\caption{
Same as Figure~\ref{fig:tevo_o}, but for a representative run in which the BH is \emph{not} trapped by the clump.
The clump parameters are $\nH = 1.5 \times 10^7\cc$ and $T = 5.6 \times 10^3\K$.  
Red solid lines show the fiducial run including ionised-bubble acceleration, while blue dashed lines correspond to a comparison run with the same initial parameters but with the ionised-bubble acceleration artificially turned off.
}
\label{fig:tevo_x}
\end{figure}
To illustrate the orbital and mass evolution, we present two representative cases: one in which the BH is captured and another in which it escapes.

Figure~\ref{fig:tevo_o} shows the successful capture case.
For $(\nH, T) = (1.8\times10^7\cc,\  5.6\times10^3\K)$, the clump mass is $\Mcl = 3.3\times10^3\Msun$.
Immediately after entry, the BH rapidly loses mechanical energy ($E < 0$), resulting in its capture.
At the same time, the BH accretes gas at a super-Eddington rate.
Accretion ceases as the clump density drops due to gas depletion, and the BH mass increases from $10^3\Msun$ to $3.9\times10^3\Msun$ over $\Delta t \sim 2\Myr$, consistent with the analytical estimate $\Dtdivsuper(\nH=1.8\times10^7\cc,\ T=6.5\times10^3\K) \simeq 1.7\Myr$ (Equation~\ref{eq:Dt_div_num}).
In this case, efficient deceleration by gas dynamical friction leads to capture.
Since the photon-trapping condition is satisfied and the stable ionized-bubble shell does not develop, the acceleration exerted by the shell is prevented, further promoting the capture.

Figure~\ref{fig:tevo_x} (red solid lines) presents a case with the same temperature but slightly lower density, $(\nH, T) = (1.5\times10^7\cc,\ 5.6\times10^3\K)$, corresponding to a clump mass of $\Mcl = 3.6\times10^3\Msun$.
In this case, the BH passes through the clump and escapes, with only negligible mass gain, $\Delta\mBH = 0.22\Msun$.
Here, the acceleration due to the shell outweighs dynamical friction and accelerates the BH rather than decelerating it.

When this shell acceleration is artificially disabled (blue dashed lines in Figure~\ref{fig:tevo_x}), the BH exits the clump with negative total energy, falls back, and eventually undergoes super-Eddington accretion.

The stronger deceleration observed at $t \approx 0$ in Figure~\ref{fig:tevo_o}, compared to the case without shell acceleration shown in Figure~\ref{fig:tevo_x}, indicates that photon trapping plays a key role in enabling BH capture.
Once photon trapping becomes effective, dynamical friction operates predominantly through the colder and denser neutral gas surrounding the ionised bubble, thereby enhancing its efficiency (see Equations~(\ref{eq:fric_II}) and~(\ref{eq:fric_I})).

In summary, BH trapping is primarily determined by the balance between ionised-bubble shell acceleration and dynamical friction.
When the photon-trapping condition is satisfied, the shell acceleration is suppressed, while dynamical friction is enhanced, enabling the clump to efficiently capture the BH.
  
\subsubsection{Clump density and temperature dependence}\label{subsubsec:clump_dependency}
\begin{figure}
	\includegraphics[bb=0 0 500 300,width=10cm,scale=0.2]{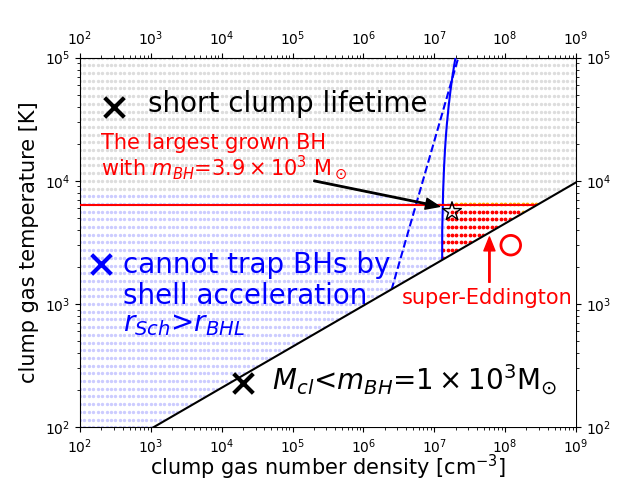}
\caption{
Summary of the fiducial runs in the clump density-temperature plane. 
Each dot represents an individual simulation.  
\textit{Red points}: the BH is trapped and undergoes super-Eddington accretion.  
\textit{Light-blue points}: the BH is not trapped.  
\textit{Grey points}: the run terminates when the clump reaches the end of its lifetime while the BH is still traversing it.  
The \textbf{black star} marks the run with the greatest BH growth, yielding $\mBHfin = 3.9 \times 10^{3}\Msun$.  
Solid and dashed curves indicate analytically derived boundaries for the BH-clump-capture model:  
- Black solid: \textit{mass doubling condition} (Equation~(\ref{eq:doubling_condition}));  
- Red solid: \textit{clump lifetime condition} (Equations~(\ref{eq:lifetime_condition_easy}) and~(\ref{eq:lifetime_condition}));  
- Blue solid: \textit{BH-trapping condition} (Equation~(\ref{eq:hyper-Eddington_condition_BHL}));  
- Blue dashed: \textit{hyper-Eddington accretion condition} evaluated at zero relative velocity (Equation~(\ref{eq:hyper-Eddington_condition})).
}
    \label{fig:rho-T_m3_v0}
\end{figure}

We now summarise the simulation results over the full $(\nH, T)$ parameter space explored in the fiducial setup.  
The outcomes of all runs are presented in Figure~\ref{fig:rho-T_m3_v0}.

As shown in the figure, BHs that achieve super-Eddington accretion are confined to a narrow region with $\nH \simeq 10^{7\ft8}\cc$ and $T \simeq (2\ft6)\times10^{3}\K$.
Whether or not super-Eddington accretion occurs can be understood analytically based on three conditions.
The first two are the \textit{mass doubling condition} ($\Mcl \geq \mBH$), which is satisfied above the black solid line, and the \textit{clump lifetime condition} ($\tcool > \Dtdivsuper$), satisfied below the red solid line.
The third is the \textit{BH-trapping condition}, which determines whether the BH is dynamically captured by the clump and is shown by the blue solid line.
Virtually all runs that meet all three analytic conditions result in super-Eddington accretion.

The blue dashed line shows the \textit{hyper-Eddington condition} (Equation~(\ref{eq:hyper-Eddington_condition})) evaluated at zero relative velocity.
In general, the BH-trapping condition is more stringent than the hyper-Eddington condition; its derivation is presented later in this section.

The largest BH growth is realised at $(\nH, T) = (1.8\times10^{7}\cc, 5.6\times10^{3}\K)$, where the BH mass increases from $\mBH = 10^{3}\Msun$ to $3.9\times10^{3}\Msun$.
This parameter set corresponds to the lowest $\nH$ and highest $T$ among the successful runs, yielding the largest Jeans mass.
Even in this most favourable case, the BH grows only by a factor of $\sim4$, suggesting that the order-of-magnitude growth reported by \citet{Lupi+2016} is not attainable for seed BHs with $\mBH = 10^{3}\Msun$.
The dependence on the BH mass is investigated in Section~\ref{subsubsec:BH_mass_dependency}.

\vspace{1em}

Whether a BH becomes trapped and enters the super-Eddington accretion phase is determined by the competition between gas dynamical friction and the acceleration exerted by the ionised-bubble shell at the moment the BH enters the clump (Section~\ref{subsubsec:representative_runs}).

We first demonstrate that, when the photon-trapping condition is not satisfied, the acceleration exerted by the ionised bubble dominates over gas dynamical friction.
The velocity of the BH when it enters the clump is given by
\bea
\vesc &=& \sqrt{\fr{2G(\Mcl + \mBH)}{\Rcl}} = 2.6\,\cs \sqrt{1 + \fr{\mBH}{\Mcl}}.
\eea
The sound velocity in the ionised region is $\csII \simeq 20\kms$, so the BH enters with a velocity in the range $\cs < \vesc < \csII$, provided that the clump temperature satisfies 
$T \lsim 6.5\times10^{3}\K$, hence $\cs\lsim 6.6\kms$.
This corresponds to a D-type ionisation front, under which ionised-bubble shell acceleration becomes significant.

Assuming $\vesc^{2} \gg \cs^{2}$, the two competing forces scale as
\bea
\fDF &\sim& 2\pi G \mBH \rho \rBHL \norm{\vesc^{2}}{\csII^{2}}{2},\nn\\
\fshell &\simeq& \pi G \mBH \rho \rStr \!\left(1 - \frac{\vesc}{2\csII}\right),\nn\\
\frac{\fshell}{\fDF} &\sim& \frac{\rStr}{\rBHL}\,
        \frac{1 - \dfrac{\vesc}{2\csII}}{\norm{\vesc}{\csII}{4}}.
\eea
Note that dynamical friction force is evaluated with the ionized gas.
Because $\vesc \lsim \csII$, the shell acceleration force exceeds the dynamical-friction force as long as the photon-trapping condition is violated ($\rStr > \rBHL$), thereby preventing BH capture by the clump.
The proportionality $\fshell/\fDF\propto\rStr/\rBHL$ can be interpreted as arising from the fact that dynamical friction is exerted by gas contained within the BHL radius $\rBHL$, whereas the ionised-bubble acceleration is generated by gas swept up to the Strömgren radius $\rStr$.

As discussed in Section~\ref{subsubsec:representative_runs}, the BH-trapping condition can be approximately reduced to the photon-trapping criterion evaluated at $v = \vesc$:
\bea
\rStr(\vesc) < \rBHL(\vesc).
\eea
This yields the following threshold for $\nH$:
\bea
\nH &>& 4.9 \times 10^{6}\cc \,\norm{T}{10^{4}\K}{1/2}\mBHsan^{-1} \nn\\
    &&\times \sqrt{1+6.7\left(1+\frac{\mBH}{\Mcl(\nH,T)}\right)}
       \,\norm{\TII}{4\times10^{4}\K}{1.4},
       \label{eq:hyper-Eddington_condition_BHL}
\eea
where $\Mcl(\nH,T)$ denotes the clump mass as a function of density and temperature.
This condition is shown as the blue solid line in Figure~\ref{fig:rho-T_m3_v0}, with $\mBH = 10^{3}\Msun$ and $\TII = 4\times10^{4}\K$, the equilibrium temperature of the ionised bubble in our simulations.

\subsection{Orbital parameter dependence}\label{subsubsec:v_b_dependency}

\begin{figure*}
    \centering
    \includegraphics[bb=0 0 1050 330,width=20cm]{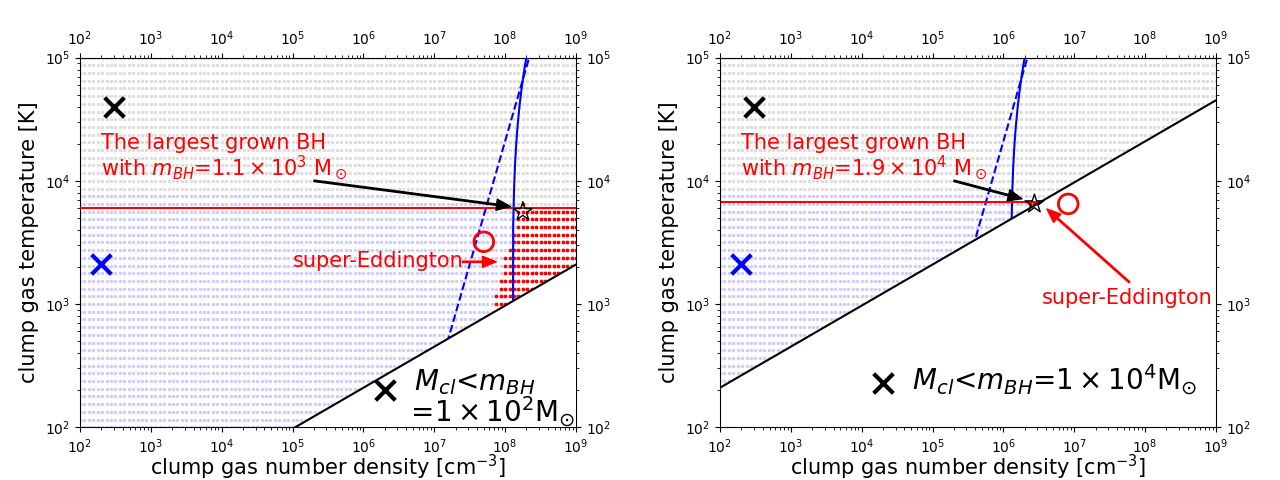}
    \caption{
    Same as Figure~\ref{fig:rho-T_m3_v0}, but for simulations with different initial BH masses: $\mBH = 10^{2}\Msun$ (left) and $\mBH = 10^{4}\Msun$ (right).
    The orbital parameters are fixed at $\vin \to 0$ and $b \to 0$ as in the fiducial setup.
    The allowed region for super-Eddington accretion shifts to higher densities and expands in the left panel ($\mBH = 10^{2}\Msun$), while it moves to lower densities and narrows significantly in the right panel ($\mBH = 10^{4}\Msun$).
    }
    \label{fig:rho-T_m2_4_v0}
\end{figure*}
In this section, we examine how the simulation outcomes change when the orbital parameters, i.e., the initial relative velocity $\vin$ and the impact parameter $b$, are varied from their fiducial, idealised values.

We begin with the initial velocity. A larger $\vin$ endows the BH with greater mechanical energy ($E > 0$) at entry, which must be dissipated by gas dynamical friction during its passage through the clump; otherwise, the BH remains unbound and exits after the encounter. We perform simulations with $\vin = 10~\kms$, representative of the virial velocity of atomic-cooling haloes, while maintaining $b \to 0$. Although the results are not shown in detail, they reveal only a modest reduction in the parameter space allowing super-Eddington accretion, by a factor of $\sim 1.3$ toward higher densities, relative to the $\vin \to 0$ runs (Figure~\ref{fig:rho-T_m3_v0}). This weak dependence arises because the entry velocity
\bea
v^{2}=\vesc^{2}+\vin^{2}\simeq6.7\left(1+\frac{\mBH}{\Mcl}\right)\cs^{2}+\vin^{2}
\eea
is dominated by the gravitational acceleration from the clump, with the contribution of $\vin\simeq\vvir$ remaining sub-dominant for $T\sim6.5\times10^{3}\K$. Despite the larger initial energy $E>0$, dynamical friction still removes sufficient energy for the BH to become bound, provided photon-trapping occurs.

Next, we consider the impact parameter. As $b$ increases, the BH passes through the outer regions of the clump, reducing both the path length through dense gas and the effectiveness of dynamical friction. For a given set of parameters $(\mBH, \nH, T, \vin)$, we define the maximum impact parameter $\bmax$ that still results in trapping, and compute the corresponding trapping cross section as $\sigmatr = \pi \bmax^{2}$. This quantity will be used in Section~\ref{subsec:Statistics}. For $\vin = 10~\kms$ and $b = \Rcl(\nH, T)$, the simulation outcomes are nearly identical to those of the $b \to 0$ case, indicating that trapping remains effective for $b \lesssim \Rcl$. In contrast, with $\vin = 10~\kms$ and $b = 2\Rcl$, the BH does not enter the clump at all for $T \lesssim 6 \times 10^{3},\K$, and no trapping occurs.

This follows from the condition on the pericentre:
\bea
\rperi=b\left(1+\frac{G\Mtot}{\Rcl\vin^{2}}\right)^{-1/2}<\Rcl,
\eea
which implies
\bea
\frac{b}{\Rcl}<\frac{\bmax}{\Rcl}=\left(1+\frac{G\Mtot}{\Rcl\vin^{2}}\right)^{1/2}. \label{eq:condition_b}
\eea
If this inequality is not satisfied, the BH merely passes by the clump without entering it. Accordingly, the trapping cross-section is bounded by
\bea
\sigmatr(\nH,T,\mBH,\vin)\lsim\pi\Rcl^{2}\left(1+\frac{G(\Mcl+\mBH)}{\Rcl\vin^{2}}\right),\label{eq:cross-section}
\eea
which evaluates to $\sigmatr\lsim2.8\pi\Rcl^{2}$ for $T\simeq6.5\times10^{3}\K$. 
Note, however, that trapping is not guaranteed even when $\rperi<\Rcl$. 
Nevertheless, we regard Equation~(\ref{eq:condition_b}) as a reasonable approximation within the parameter range of interest ($T\lsim6\times10^{3}\K$, $\vin\sim10\kms$). 
Given that the BH remains trapped at $b=\Rcl$, the effective trapping cross-section lies within the range 
$\sigmatr=(1\ft3)\x\pi\Rcl^2$.  

\subsection{BH mass dependence}\label{subsubsec:BH_mass_dependency}

We now examine how BH growth depends on the initial BH mass by comparing the results with the fiducial case of $\mBH = 10^3\Msun$.  
Figure~\ref{fig:rho-T_m2_4_v0} presents simulation outcomes for $\mBH = 10^2\Msun$ (left) and $10^4\Msun$ (right), adopting the idealised orbital parameters $\vin \to 0$ and $b \to 0$.  
In both panels, the qualitative behaviour remains similar to that described in Section~\ref{subsubsec:clump_dependency}: super-Eddington accretion is realized only when all three analytical criteria are simultaneously satisfied.

For $\mBH = 10^2\Msun$ (left panel), the region permitting super-Eddington growth shifts toward higher densities and expands across a broader area of the $(\nH, T)$ plane.
The maximum final mass achieved in this case is $1.1 \times 10^{3}\Msun$, corresponding to a growth factor of approximately ten within a single encounter.
In contrast, for $\mBH = 10^4\Msun$ (right panel), the viable region moves to lower densities but becomes extremely narrow: only a single run attains super-Eddington accretion.
The final mass in that case is $1.9 \times 10^{4}\Msun$, indicating that growth by at most a factor of two is feasible, and only within clumps with masses $\Mcl \sim 10^{4}\Msun$.

\begin{figure}
    \includegraphics[bb=0 0 500 420,width=10cm]{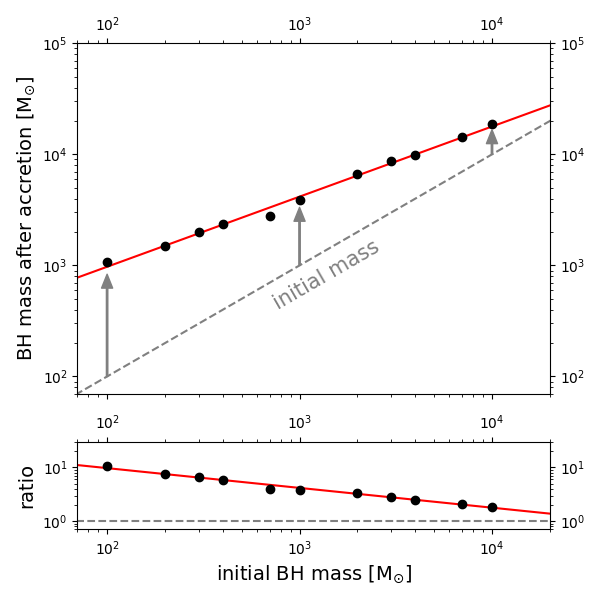}
    \caption{
    \textbf{Top:} Maximum final BH mass $\mBHfin$ attained via super-Eddington accretion for each initial BH mass $\mBH$.
    Black points represent the simulation results; the red curve shows the best-fit relation (Equation~\ref{eq:saidai}), and the grey dashed line shows the identity relation ($\mBHfin = \mBH$), corresponding to no mass growth.
    \textbf{Bottom:} Corresponding BH growth factor $\mBHfin/\mBH$ as a function of initial mass.
    The red curve is the best-fit relation given in Equation~\ref{eq:seityouritsu}, while the grey dashed line marks unity (no growth).
    }
    \label{fig:m0_combined}
\end{figure}

To quantify this trend more systematically, we extend our analysis to a broader range of initial BH masses.
Figure~\ref{fig:m0_combined} summarises the dependence on $\mBH$.
For seeds with even higher initial masses ($\mBH \geq 2 \times 10^{4}\Msun$), none of the runs enter the super-Eddington regime within the explored $(\nH, T)$ parameter space.
  
The upper panel of Figure~\ref{fig:m0_combined} shows the maximum final BH mass $\mBHfin$ achieved in each simulation.  
A fit to the data yields
\bea
\mBHfin &\leq& 4.2 \times 10^{3}\Msun \; \mBHsan^{0.63}, \label{eq:saidai}
\eea
from which the corresponding upper limit on the growth factor becomes
\bea
\frac{\mBHfin}{\mBH} &\leq& 4.2 \; \mBHsan^{-0.37}. \label{eq:seityouritsu}
\eea
This confirms that for seed masses $\mBH \gtrsim 10^{4}\Msun$, the growth factor satisfies $\mBHfin/\mBH \lesssim 2$; even under the most favourable conditions ($\nH, T, \vin, b$), the mass increases by no more than a factor of two.  
We therefore conclude that the BH-clump-capture mechanism is efficient only for seed masses below $\sim 10^{4}\Msun$.

We also investigate the allowed region in the $(\nH,\mBH)$ parameter space at fixed temperature $T = 6.5 \times 10^{3}\K$, corresponding to the upper limit imposed by the lifetime condition (Equation~(\ref{eq:lifetime_condition_easy})).
The results are shown in the lower panel of Figure~\ref{fig:m0_combined}.
They indicate that BH trapping and subsequent super-Eddington accretion require approximately
\bea
\nH \, \mBH &\gtrsim& 2 \times 10^{10}\cc\Msun, \label{eq:nH-m_rel_1}
\eea
which can be more precisely fitted by
\bea
\nH &\geq& 2.1 \times 10^{7}\cc \; \mBHsan^{-0.88}. \label{eq:nH-m_rel_2}
\eea
This condition is about an order of magnitude more stringent than the criterion proposed by \citet{Inayoshi+2016}.
Furthermore, clumps that are too dense, satisfying $\nH > 3.3 \times 10^8\cc~\mBHsan^{-2}$, are excluded by the mass doubling condition (Equation~(\ref{eq:doubling_condition})), since in such cases the clump mass $\Mcl$ is smaller than the BH mass $\mBH$ and therefore insufficient for growth.

The declining growth factor with increasing $\mBH$ originates from the opposing dependencies of the doubling and BH-trapping conditions on BH mass.
As $\mBH$ increases, the parameter space satisfying both conditions narrows, reducing the potential for mass growth.

In summary, BH growth via the BH-clump-capture mechanism is more efficient for lighter seeds.
An initial BH with $\mBH = 10^{2}\Msun$ can grow up to $\sim 10^3\Msun$ in a single encounter.
However, the growth factor decreases with increasing $\mBH$, and the mechanism becomes ineffective once the BH mass approaches $\sim 10^{4}\Msun$.
This implies that the clump-capture process alone cannot drive light seeds to $\sim 10^{5}\Msun$; additional mechanisms must operate to achieve such growth.

\begin{figure}
    \includegraphics[bb=0 0 520 320,width=10cm]{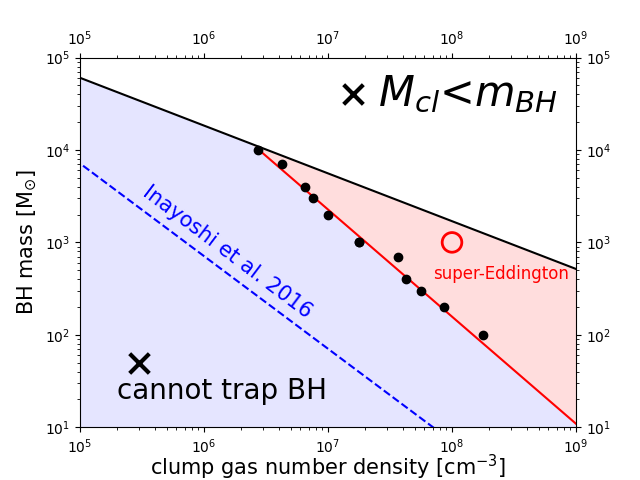}
    \caption{
    Parameter space in the $(\nH,\mBH)$ plane allowing super-Eddington accretion in the BH-clump-capture model.  
    Black points denote the runs with the largest mass gain for each $\mBH$; the red curve is a fit to these points.  
    The black solid line shows the mass doubling condition ($\Mcl > \mBH$) at the maximum temperature allowed by the lifetime condition, $T = 6.5 \times 10^{3}\K$.  
    Super-Eddington accretion and trapping occur only in the region bounded by the red and black curves.  
    The blue dashed line indicates the criterion of \citet{Inayoshi+2016}.
    }
    \label{fig:nHmax_m0}
\end{figure}

\section{DISCUSSION}\label{sec:discussion}
Section~\ref{subsec:implication} explores the broader implications of our results.  
Section~\ref{subsec:Statistics} evaluates the statistical abundance of clumps and the resulting long-term BH growth rate.  
The remaining subsections examine physical processes omitted from our model, including turbulence (Section~\ref{subsec:turbulence}), mechanical feedback (Section~\ref{subsec:mechanical}), and Ly$\alpha$ radiation pressure (Section~\ref{subsec:Lyman_alpha}).

\subsection{Implications of ionised bubble acceleration}\label{subsec:implication}

Here we discuss the broader implications of our results, focusing on the acceleration exerted by the ionised shell.

If the ambient gas is neutral (or molecular) and not sufficiently dense to satisfy the photon-trapping criterion (see Equation~(\ref{eq:nH-m_rel_2})), a D-type ionisation front forms around the BH and, once a stable shell is established, exerts a forward acceleration \citep{Sugimura&Ricotti+2020, Toyouchi+2020} whenever
\bea
    \cs < v &\lsim& \csII + \sqrt{\csII^{2}-\cs^{2}} \equiv \vR ,
\eea
where $\csII$ is the sound velocity inside the ionised bubble, $\cs$ that of the surrounding neutral gas, and $v$ the BH-gas relative velocity (see Appendix~\ref{Appx:Park-Ricotti_model}).  
Section~\ref{subsubsec:clump_dependency} showed that, under such conditions, shell acceleration exceeds dynamical friction.  
Hence $v$ decreases for $v<\cs$ (or $v>\vR$) owing to friction, but increases for $\cs < v < \vR$; a BH with $v>\cs$ therefore evolves towards the terminal velocity $v\rightarrow\vR$.

This mechanism could undermine all super-Eddington growth scenarios based on light seeds, not only the BH-clump-capture model.  
Whenever a gas reservoir with $\Mgas>\mBH$ sets the Bondi-scale boundary conditions, shell acceleration ejects the BH unless the BH-trapping condition is met.  
Because the BH-trapping and mass doubling conditions conflict at large masses, super-Eddington growth ceases once the BH reaches $\mBH\sim10^{4}\Msun$.

Shell acceleration may also impede the orbital decay of BHs towards galactic centres.  
Some SMBH formation scenarios assume that dynamical friction carries BHs inward, enabling efficient accretion.  
If $\cs < v < \vR$, however, the BH is accelerated outward rather than decelerated.  
In such haloes, the BH may fail to settle in the dense nucleus, so the confining effect of nuclear discs or clusters would be lost, rendering super-Eddington growth even more challenging.

The difficulty is most serious for haloes with $\vvir < \vR \simeq 50\kms (\TII/4\times10^{4}\K)^{1/2}$, namely
\be
\vvir=\sqrt{\frac{G\Mhalo}{\Rvir}}<50\kms
\quad\LRar\quad
\Mhalo<10^{9}\Msun\left(\frac{1+z}{20}\right)^{-3/2}.
\ee
Such low-mass haloes exist only for a few $\times10^{2}\,$Myr after the Big Bang; during this epoch the gas becomes metal-enriched, further suppressing the formation of massive, dense clumps and, hence, super-Eddington accretion\footnote{
For $\mBH=10^{4}\Msun$, shell acceleration has been discussed by \citet{Toyouchi+2020, Ogata+2024}.  Those studies, however, neglected fragmentation and clump stability.  \citet{Toyouchi+2020} concentrated on the competition between acceleration and friction without assessing BH growth, while \citet{Ogata+2024} extrapolated accretion rates over $\Delta t>16\,\Myr$ for gas with $\nH\simeq10^{4\ft6}\cc$ and $T=180\K$ and suggested the rapid BH growth.  Our analysis suggests that such gas fragments into $\Mcl\sim8\times10^{2}\Msun$ clumps with clump-crossing time $\tcross\sim0.5\,\Myr$, implying that their extrapolation requires $\sim10^{1\ft2}$ successive encounters and is therefore optimistic when compared with the collision rates in Section~\ref{subsec:Statistics}.
}.

To avoid these obstacles, a seed must form inside, and remain within, gas dense enough for photon trapping.  
The required scale, $\sim0.2\pc \, \mBHsan^{1/2}\simeq10^{-3}\,r_{\rm vir}$, implies that seed formation, gas inflow, and super-Eddington growth must proceed as a continuous sequence.
A plausible candidate for the process is cold streams feeding the halo centre \citep{Inayoshi&Omukai2012, Latif+2022, Kiyuna+2023, Kiyuna+2024}.

Our discussion ignores stellar dynamical friction, gas inhomogeneity, and the finite time for the ionisation front to expand.  
In regions where the stellar mass is comparable to or higher than the gas mass, stars may be the dominant source of dynamical friction.  
If the ionised bubble approaches galactic scales, other gravitational forces can dominate, diminishing the shell acceleration.  
Highly anisotropic gas distributions would also invalidate spherical assumptions, requiring more sophisticated modeling.

Directly resolving shell acceleration requires radiation-hydrodynamic simulations with spatial resolution $\sim0.01 \ft 0.1 \pc \, \mBHsan^{1/3}(\nH/10^{4}\cc)^{-2/3}$ \citep{Toyouchi+2020}, beyond current cosmological capabilities for small seeds.  
Future work should therefore test the effect with dedicated simulations and incorporate a suitable sub-grid model to evaluate its impact on early BH growth.

\subsection{Statistical evaluation of BH-clump capture}\label{subsec:Statistics}
Throughout this paper we have treated the clump density $\nH$ and temperature $T$ as control parameters, leaving open the question of how often such clumps actually form and collide with BHs.  
As a reference, we here evaluate the BH-clump collision rate and the resulting statistical BH growth rate.

As a representative early-universe environment we adopt atomic-cooling haloes with virial temperature $\Tvir \sim 10^{4}\K$.  
We assume a seed BH moving at a fixed velocity within a sub-volume of the halo and introduce the following simplifying assumptions:  
\begin{itemize}
    \item The gas is taken to be isothermal at $T = 6.5 \times 10^{3}\K$, corresponding to the maximum temperature allowed by the clump-lifetime condition (Equation~(\ref{eq:lifetime_condition_easy})).  
          This is optimistic; in reality clumps would exhibit a different temperature or a range of temperatures.
    \item Whenever a BH is trapped by a clump, the entire clump mass accretes on to the BH.  
          In our simulations (Section~\ref{subsubsec:representative_runs}), the accreted fraction was $\simeq 3/4$; the difference is only a factor of order unity and is neglected.
    \item The time from trapping to the end of accretion is $\sim\,$Myr, but provided the mean collision time interval is much longer we may treat accretion as instantaneous.
    \item The relative velocity between BH and clump is fixed at the virial velocity, $\vin = 10\kms$.
\end{itemize}

For clarity, we first suppose that all clumps share a single density $\nH = 1.8 \times 10^{7}\cc$, the value that yielded the largest BH growth for $\mBH = 10^{3}\Msun$ (Section~\ref{subsec:fiducial}).  
Denoting the clump number density by $\nclump$, the collision rate $\Gamma$ and the statistical BH accretion rate $\langle\mdotBH\rangle$ are
\bea
\Gamma &=& \nclump\,\sigmatr\,\vin, \label{eq:gamma_1}\\
\langle\mdotBH\rangle &=& \Mcl\,\Gamma, \label{eq:mdot_toukei_1}
\eea
where the trapping cross-section is approximated as $\sigmatr \simeq 2.8\,\pi\Rcl^{2}$ (Equation~(\ref{eq:cross-section})).  
Hence
\bea
\langle\mdotBH\rangle &\simeq& 2 \times 10^{-5}\Msun\yr^{-1}\!
\left(\frac{\nclump}{3 \times 10^{-3}\pc^{-3}}\right)
\left(\frac{\vin}{10\kms}\right) \nonumber\\
&&\times\left(\frac{\nH}{1.8 \times 10^{7}\cc}\right)^{-3/2}
\left(\frac{T}{6.5 \times 10^{3}\K}\right)^{5/2}.
\eea
Achieving a rate above the Eddington value,
$\mdotEdd = 2 \times 10^{-5}\Msun\yr^{-1}\,\mBHsan$, therefore requires $\nclump \gtrsim 3 \times 10^{-3}\pc^{-3}$.  
By comparison, \citet{Lupi+2016} found only $\sim\!10$ clumps in their CND of radius $\rdisc \simeq 50\pc$; meeting the requirement would demand $\sim\!100$ clumps.

The above estimate considered a single density.  
In reality clump density follow a distribution.  
We adopt the log-normal density PDF for turbulence-dominated gas proposed by \citet{Hennebelle_Chabrier2008}.  
Although developed for $T = 10\K$ star-forming clouds, we assume it applies to Ly$\alpha$-cooled gas at $T \simeq 6.5 \times 10^{3}\K$.  
For a mean density $\nHbar$, the volume fraction at density $\nH$ is
\bea
\frac{1}{V}\frac{\rmd V}{\rmd\ln\nH} &=& \frac{1}{\sqrt{2\pi\sigma^{2}}}
\exponentHS,\\
\sigma^{2} &\simeq& \ln(1+\bturb\Machturb^{2}),
\eea
with $\bturb = 0.25$ and $\Machturb = 2$, calibrated with \citet{WiseAbel2007}, giving $\sigma^{2}=0.69$.  
The clump number-density distribution is then
\be
\frac{\rmd\nclump}{\rmd\ln\nH} =
\frac{\rho}{\Mcl(\nH,T)}
\frac{1}{\sqrt{2\pi\sigma^{2}}}
\exponentHS.
\label{eq:kosuubunnpu}
\ee

The contribution to BH growth from clumps between $\nH$ and $\nH(1+\rmd\ln\nH)$ is
\bea
\rmd\langle\mdotBH\rangle =
\Mcl \frac{\rmd\nclump}{\rmd\ln\nH}\,\sigmatr\,\vin\,\rmd\ln\nH. \label{eq:mdot_toukei_2}
\eea
Substituting Equation~(\ref{eq:kosuubunnpu}) and integrating over $\nHmin < \nH < \nHmax$ that satisfy the trapping condition yields
\bea
\langle\mdotBH\rangle &\simeq&
\rhomin\,\sigmatr(\nHmin)\,\vin\nn\\
&\x& \int_{\nHmin}^{\nHmax}
\frac{\rmd\ln\nH}{\sqrt{2\pi\sigma^{2}}}
\exponentHS \nonumber\\
&\equiv& \rhomin\,\sigmatr(\nHmin)\,\vin~\fcl \nonumber\\
&\simeq& 0.75\Msun\yr^{-1}
\fcl\left(\frac{T}{6.5\times10^{3}\K}\right)
\left(\frac{\vin}{10\kms}\right)\nn\\
&&
\eea
Here $\sigmatr\propto\nH^{-1}$, so $\rho\sigmatr$ can be taken outside the integral.  
The fraction of clumps satisfying the trapping condition is
\bea
\fcl \simeq \frac{1}{2}\,{\rm erfc}(y),\qquad
y = \frac{\ln(\nHmin/\nHbar)+\sigma^{2}/2}{\sqrt{2}\sigma}.
\eea
Requiring $\langle\mdotBH\rangle \ge \mdotEdd$ gives
$\fcl \gtrsim 2.7\times10^{-5}(\mBH/10^{3}\Msun)^{-1}$, corresponding to $y < 2.86$ and hence
\be
\nHbar \;>\; 8.9 \times 10^{5}\cc
\left(\frac{\nHmin}{1.8 \times 10^{7}\cc}\right).
\ee
Thus sustained super-Eddington growth, $\langle\mdotBH\rangle \ge \mdotEdd$, via clump collisions requires the seed BH to lie in gas of mean density $\nHbar \gtrsim 8.9 \times 10^{5}\cc$.  
Such densities are confined to regions of characteristic scale $\sim 0.2\pc\,(T/6.5 \times 10^{3}\K)^{1/2}$, about three orders of magnitude smaller than the halo virial radius $\rvir \gtrsim 300\pc$ for $\Mhalo = 10^{7}\Msun$ at $z \simeq 19$.  
This scale is comparable to the radius of a clump capable of sustaining super-Eddington accretion, $\Rcl \sim 0.1\pc(\nH/1.2 \times 10^{7}\cc)^{-1/2}(T/6.5 \times 10^{3}\K)^{1/2}$.  
This stringent requirement re-invokes the fine-tuning problem highlighted in Section~\ref{sec:introduction}: how can a light seed encounter such a dense CND without relying on its own gravity?

These estimates are, of course, crude.  
They neglect the fragmentation mode due to chemo-thermal instability, angular momentum, and others.  
A more detailed evaluation requires cosmological radiation-hydrodynamic simulations, which we leave to future work.

\subsection{Influence of turbulence}\label{subsec:turbulence}
In this study we have ignored the potential influence of turbulence when analysing the internal structure of clumps and the gas accretion on to the BH.  By contrast, turbulence is regarded as essential in local star-formation theory: clumps are thought to arise through turbulent compression and to be sustained mainly by the effective pressure of turbulent motions \citep[e.g.][]{Hennebelle_Chabrier2008}.
This modifies, for example, the Jeans mass in Equation~(\ref{eq:Jeans_Mass}) to  
\bea
\aJ \;\to\; \aJ\left(1 + \frac{\mathcal{M_{\rm turb}}^{2}}{3}\right)^{3/2},
\eea
where $\Machturb \equiv \vturb/\cs$ is the turbulent Mach number.
We therefore consider how important turbulence may be for the clumps in this study.

Turbulence decays unless it is continuously driven, on an eddy-turnover time of only a few to ten free-fall times $\tff$.  By contrast, the BH mass growth considered here requires a duration more than an order of magnitude longer than $\tff$ (Equation~(\ref{eq:BH_div})); in our fiducial run (Section~\ref{subsubsec:representative_runs}) the growth phase lasted about $170\tff$.

Even if a clump were initially supported by turbulence, that support would decay rapidly, so the clump would collapse or fragment on a timescale far shorter than that required for BH growth.  We therefore conclude that, although turbulent compression may form clumps, turbulence-support for clumps does not contribute in our scenario, and omitting turbulence from our analysis is justified.

\subsection{Mechanical feedback from slim disc outflows}\label{subsec:mechanical}

In this study, we have so far taken into account only radiative feedback from the accretion disc when evaluating BH growth via super-Eddington accretion.  
Recent simulations \citep[e.g.][]{Hu+2022, Toyouchi+2024, Fragile+2025} have shown that powerful outflows are launched from slim discs.  
Such outflows may inject significant momentum and energy into the surrounding clump, potentially disrupting it and preventing it from surviving over extended periods.

The impact of the outflow on the clump strongly depends on its geometrical structure, particularly the degree of collimation, and is therefore not straightforward to evaluate.  
If the outflow is less collimated, it may collide with the clump isotropically, injecting momentum and energy efficiently, which promotes disruption.  
Conversely, a highly collimated outflow may escape without significantly interacting with the clump.  
\citet{Toyouchi+2024} suggest that, in practice, the situation lies between these two extremes.

Outflows from the slim disc are launched at each radius $r$ with a velocity roughly given by $v \sim \sqrt{G\mBH/r}$.  
In general, outflows from the outer region are less collimated and slower, while those from the inner region are more collimated and faster.  
Accordingly, the mass outflow rate follows $\dot{M}_{\rm out}(r) \propto r^{0.5}$, dominated by the slow, outer outflows,  
whereas the energy injection rate, $\dot{M}_{\rm out} v^2 \propto r^{-0.5}$, is dominated by the fast, inner outflows.

In this paper, we adopt an optimistic assumption: the outer, slow outflows do not escape the clump and eventually fall back to the disc, while the inner, fast outflows escape through the clump with negligible energy injection.  
Under this assumption, we regard both mass loss from the clump and energy injection into the clump negligible.  
While this assumption should be tested with future radiation hydrodynamic simulations, such an investigation is beyond the scope of the present work.

Nevertheless, even if the energy injection is neglected, momentum injection from the outflow may still unbind the clump gravitationally and lead to its disruption \citep{King2003,Kido+2025}.  
The momentum injection rate, $\dot{P} \propto \dot{M} v \propto r^0$, implies roughly equal contributions from all disc radii.  
Among these, slow outflows from the outer regions are more likely to shock with the clump, and thus at least some fraction of the outflow momentum will be deposited into it.

If the injected momentum raises the clump's kinetic energy above its gravitational binding energy, the clump can no longer remain bound and will expand and eventually disrupt.  
To evaluate this possibility, we estimate the timescale required for such disruption and compare it with the timescale over which the BH mass grows significantly.  

The momentum corresponding to the gravitational binding of the clump is given by:
\bea
\Pcl &\simeq& \Mcl \sqrt{\fr{G \Mcl}{\Rcl}}.
\eea
The momentum injection rate from the slim disc is estimated as \citep{Hu+2022}:
\bea
\dot{P} &=& \zeta \mdotBH c,
\eea
where we adopt $\zeta = 0.03$, corresponding to the case where the entire outflow momentum injects into the clump.
The timescale for the clump to be disrupted by the injected momentum is then:
\bea
\Delta \tp &\simeq& \fr{\Pcl}{\dot{P}} = \zeta^{-1} \left( \fr{\Mcl}{\mBH} \right) \left( \fr{\mBH}{\mdotBH} \right) \sqrt{\fr{G \Mcl}{c^2 \Rcl}} \nn \\
&\simeq& \zeta^{-1} \left( \fr{\Mcl}{\mBH} \right) \sqrt{\fr{G \Mcl}{c^2 \Rcl}} \x  \Delta t_{\rm div},
\eea
where we have used Equations~(\ref{eq:Bondi_acc_rate}) and (\ref{eq:BH_div}).

To ensure that the BH can grow significantly before the clump is disrupted, the condition
\bea
\Delta \tp &>& \Dtdivsuper \nn \\
\Leftrightarrow T &\gsim& 2.4 \x  10^5\K \norm{\nH}{2 \x  10^7\cc}{1/4} \norm{\mBH}{10^3\Msun}{1/2} \norm{\zeta}{0.03}{1/2}\nn\\
&&
\eea
must be satisfied.

However, this condition cannot be met within the clump lifetime constraint (Equation~(\ref{eq:lifetime_condition})), which limits the gas temperature to $T \lsim 6.5 \x  10^3\K$.  
Therefore, for $\mBH > 1\Msun$, clumps are likely to be disrupted by momentum injection before the BH can undergo substantial growth.

The value $\zeta = 0.03$ corresponds to the case where all of the outflow momentum is deposited into the clump.  
In practice, if inner fast outflows escape without interacting significantly with the clump, the effective value of $\zeta$ can be smaller by a factor of a few.  
The maximum allowable value of $\zeta$ that satisfies the condition $\Delta \tp > \Dtdivsuper$ is:
\bea
\zeta < 6 \x  10^{-5} \left( \fr{T}{6.5 \x  10^4\K} \right)^2,
\eea
which is 2-3 orders of magnitude smaller than $\zeta = 0.03$.  
This would require nearly all outflows from the disc to pass through the clump without depositing momentum, a condition unlikely to be realized based on the results of \citet{Toyouchi+2024}.

For this reason, although we have adopted an optimistic assumption that neglects mechanical feedback, it remains entirely possible that momentum injection from the outflow disrupts the clump, thereby suppressing BH growth.  
A rigorous evaluation of this effect will require detailed radiation hydrodynamic simulations in future work.

\subsection{Ly$\alpha$ Radiation Pressure}\label{subsec:Lyman_alpha}

In this study, we consider only Thomson scattering by free electrons as the dominant opacity source in the radiative feedback from the BH accretion disc.  
This assumption is consistent with many previous studies on super-Eddington accretion \citep[e.g.][]{Inayoshi+2016}.

However, recent works have pointed out that in slim discs undergoing super-Eddington accretion, radiation pressure driven by Ly$\alpha$ line force may play a significant role in enhancing radiative feedback.  
\citet{Mushano+2024} performed numerical simulations posing a static gas surrounding the slim disc, and evaluated the strength of Ly$\alpha$ radiation pressure acting on the gas.  
Their results suggest that maintaining hyper-Eddington accretion requires a stricter condition, $\mBH \x  \nH \gsim 4 \x  10^{10}\Msun\cc$,  
which is an order of magnitude more stringent than the threshold $\mBH \x  \nH \gsim 10^9\Msun\cc$ proposed by \citet{Inayoshi+2016}.

If such enhanced radiation pressure is taken into account, the BH-trapping condition derived in this study (Equation~(\ref{eq:hyper-Eddington_condition_BHL})) may also become more stringent.
We note that the effectiveness of Ly$\alpha$ radiation pressure depends sensitively on the geometry and kinematics of the gas.  
As a result, the actual degree to which super-Eddington accretion is suppressed remains uncertain and will need to be tested through future high-resolution radiation hydrodynamics simulations.

\section{CONCLUSION}\label{sec:conclusion}

We have investigated, through a combination of analytic arguments and dedicated toy-model simulations, whether a light-seed BH can grow beyond the $\sim10^{5}\Msun$ bottleneck via single or multiple clump-encounter events that trigger super-Eddington accretion.  Compared with earlier studies \citep[e.g.][]{Lupi+2016}, which were limited by spatial resolution, our analytical formulation improves on them by consistently tracking (i) gas accretion at the BH Bondi radius, (ii) radiative feedback, and (iii) both the gas dynamical friction and the forward-acceleration effect exerted by the ionised shell.  Although simplified, our toy model already reveals a sequence of requirements for super-Eddington growth.  The principal conclusions are as follows.

First, we have shown that the mass gained during a single passage across a clump is negligible (Section~\ref{subsec:3_conditions}); a BH must remain gravitationally \emph{trapped} for a period that is at least orders of magnitude longer than the clump crossing time.  This demands three simultaneous requirements:
\begin{itemize}
  \item \textbf{Mass doubling condition (Section~\ref{sec:doubling_condition}):} the clump mass must exceed the BH mass so that the BH can, in principle, increase by a factor of two or more in a single encounter, enforcing $\Mcl>\mBH\LRar\nH<3.3\x10^8\cc~\mBHsan^{-2}(T/6.5\x10^3\K)^{3}$.
  \item \textbf{Lifetime condition (Section~\ref{sec:lifetime_condition}):} the radiative cooling time of the clump must exceed the super-Eddington growth time; for primordial gas this enforces $T\lesssim6.5\times10^{3}\,$K.
  \item \textbf{BH-trapping condition (Section~\ref{sec:trap_condition}):} dynamical friction must dominate ionised bubble acceleration, which is possible only when the photon-trapping criterion $\rBHL>\rStr$ is satisfied at the moment of entering the clump with $v=\vesc$. This condition is transformed to $\nH\geq2.1\x10^7\cc\mBH^{-0.88}$.
\end{itemize}
These constraints define a narrow \emph{sweet spot} in the density-temperature plane.  A BH with $10^{3}\Msun$ can be trapped and undergo super-Eddington accretion only for $n_{\mathrm H}\simeq10^{7\ft8}\cc$ and $T\simeq(2\ft6)\times10^{3}\,$K (Section~\ref{subsubsec:clump_dependency}).

Second, even within this sweet spot a seed with $\mBH=10^{3}\Msun$ grows at most to $4\times10^{3}\Msun$, i.e.\ by a factor of $\sim4$.  Extending the calculations over a wider range of initial masses shows that the maximal growth factor $\mBHfin/\mBH$ declines roughly as $\mBH^{-0.4}$; once BH mass reaches $\mBH\gtrsim10^{4}\Msun$, the BH-clump-capture model ceases to operate.  While the mechanism may assist the earliest growth of light seeds ($\mBH<10^{4}\Msun$), it cannot produce the $\gtrsim10^{5}\Msun$ seeds required for subsequent Eddington-limited growth (Section~\ref{subsubsec:BH_mass_dependency}).

Third, we have demonstrated that the forward-acceleration effect exerted by the ionised shell \citep{Toyouchi+2020, Ogata+2024} is generally stronger than gas dynamical friction unless the photon-trapping condition is met (Section~\ref{subsec:fiducial}).  The shell force gets the BH out of the dense gas, hindering further accretion within it.
This effect has serious influences in various situations focusing on gas dynamical friction, including other light-seed super-Eddington models, and is suggested to suppress rapid growth in them (Section~\ref{subsec:implication}).

Taken together, our results indicate that, attractive though it is in concept, the BH-clump-capture model is unlikely to bridge the gulf between light seeds and the $\sim10^{7-9}\Msun$ SMBHs observed at $z\gtrsim6$. 

Our study highlights several avenues for future work.  High-resolution radiation-hydrodynamic simulations are required to verify the ionised-shell force in realistic environments including non-uniform density structures and other radiation sources.  In addition, cosmological simulations that incorporate this effect via suitable sub-grid models will be essential for reassessing the true feasibility of super-Eddington growth and for tracing the growth history of the earliest BHs.

\section*{ACKNOWLEDGEMENTS}
The author expresses sincere gratitude to Prof. Takahiro Tanaka for his continuous interest and encouragement.
I am also deeply grateful to Takashi Hosokawa and Kazuyuki Omukai for their valuable discussions, insightful suggestions, and careful reading of the manuscript.
I would like to thank Kunihito Ioka, Kazuyuki Sugimura, Kazumi Kashiyama, Daisuke Toyouchi, Kazutaka Kimura, Ryunosuke Maeda, and Tomoya Suzuguchi for fruitful discussions and helpful comments.
This research was supported by Grants-in-Aid for Scientific Research (19H01934, 19KK0353, 22H00149) from the Japan Society for the Promotion of Science and by JST SPRING, Grant Number JPMJSP2110.

\section*{DATA AVAILABILITY}
The data underlying this article will be shared on a reasonable request of the corresponding author.



\bibliographystyle{mnras}
\bibliography{ms} 


\appendix
\section{PARK-RICOTTI MODEL FOR RADIATION FEEDBACK}
\label{Appx:Park-Ricotti_model}

Cosmological hydrodynamic simulations typically adopt simple prescriptions for BH accretion, such as the BHL accretion rate evaluated in the immediate vicinity of the BH.  
These prescriptions is believed to reproduce the dynamics provided that the hydrodynamics is resolved and radiative feedback is modelled consistently.  
In this paper, we do not solve the radiation hydrodynamics in the gas clump explicitly, and such standard prescriptions therefore is inadequate.  
We instead employ a more sophisticated accretion model that treats the coupling between accretion, radiative feedback, and the gas properties around the BH.

This section introduces the ``Park-Ricotti'' model, which determines the properties of the ionised bubble.  
We consider the BH within the ionised bubble that is embedded in a neutral ambient medium.  
We first assume a trial temperature for the ionised gas and corresponding sound velocity, as
\bea
\TII &=& 4\times10^{4}\,\K, \label{eq:TII}\\
\csII &=& \sqrt{\frac{\kB\TII}{\muII\mH}}, \label{eq:csII}
\eea
where we adopt $\muII\simeq0.6$ for fully ionised gas.  
Our model then returns the ionised-bubble density $\nHII$, the relative velocity inside the bubble $\vII$, and the BHL accretion rate evaluated within the bubble, $\mdotBHLII$, once the following parameters are specified: the BH mass $\mBH$, the ambient density $\nH$, temperature $T$, and relative velocity with respect to the BH $v$, together with the assumed bubble temperature $\TII$.  
Using these quantities, the accretion model computes the actual BH accretion rate and luminosity, from which we derive the radiative cooling and photoionisation heating rates within the bubble (Appendix~\ref{Appx:BH_accretion_model}).  
If heating exceeds cooling we raise $\TII$, whereas we lower it when cooling dominates.  
We iterate this procedure until chemo-thermal equilibrium, in which the heating and cooling rates balance, is achieved.

 Given $\TII$, the density $\nHII$ and relative velocity $\vII$ inside the bubble follow the jump conditions across the ionisation front.  
Their functional form depends on $v$, the neutral sound velocity $\cs$, and $\csII$ as follows \citep{Park&Ricotti2013,Sugimura&Ricotti+2020}.

\subsection*{R-type front: $v>\vR\equiv\csII+\sqrt{\csII^{2}-\cs^{2}}\equiv \vR$}

For an R-type front the contrasts across the front are modest:
\bea
\Delta^{(-)}(v,\cs,\csII) &=& 
\frac{\cs^{2}+v^{2}-\sqrt{(\cs^{2}+v^{2})^{2}-4\csII^{2}v^{2}}}{2\csII^{2}},\\
\rhoII &=& \Delta^{(-)}\rho,\\
\vII &=& \frac{v}{\Delta^{(-)}}.
\eea
Here $\Delta^{(-)}\rightarrow1$ as $v\rightarrow\infty$.  
The BHL accretion rate evaluated inside the bubble is
\bea
\mdotBHLII
  &=& \frac{4\pi G^{2}\mBH^{2}\rhoII}{(\csII^{2}+\vII^{2})^{3/2}}\nonumber\\
  &=& \frac{4\pi G^{2}\mBH^{2}\Delta^{(-)}\rho}
          {\left[\csII^{2}+(\Delta^{(-)})^{-2}v^{2}\right]^{3/2}}.
\eea

\subsection*{D- type front: $\cs<v<\vR$}

When $\cs<v<\vR$ a shock precedes the ionisation front, decelerating the gas to $\vII\simeq\csII$ and producing a pressure-balanced bubble:
\bea
\vII &=& \csII,\\
\rhoII &=& \frac{\cs^{2}+v^{2}}{2\csII^{2}}\rho.
\eea
The corresponding BHL accretion rate is
\bea
\mdotBHLII
  = \frac{4\pi G^{2}\mBH^{2}\rho(\cs^{2}+v^{2})}{2\sqrt{2}\,\csII^{5}}.
\eea

\subsection*{Sub-sonic case: $v<\cs$}

For sub-sonic motion the bulk velocity inside the bubble is not dynamically important.  
Following \citet{Sugimura&Ricotti+2020} we adopt
\bea
\vII &=& \left(\frac{\csII}{\cs}\right)v,\\
\rhoII &=& \frac{\cs^{2}}{\csII^{2}}\rho,
\eea
which yields
\bea
\mdotBHLII
  = \frac{4\pi G^{2}\mBH^{2}\rho}{(\cs^{2}+v^{2})^{3/2}}
    \left(\frac{\cs}{\csII}\right)^{5}.
\eea

\section{BH ACCRETION MODEL}
\label{Appx:BH_accretion_model}

This section outlines our prescription for BH accretion within the ionised bubble that is embedded in a neutral ambient medium.  
Given the BH mass $\mBH$ and the gas density, temperature, and relative velocity inside and outside the bubble, $\left(\nHII,\TII,\vII\right)$ and $\left(\nH,T,v\right)$, respectively, the model assess the photon-trapping condition and adopts the appropriate accretion regime, sub-Eddington, Eddington-limited, or hyper-Eddington, returning the accretion rate $\mBHdot$ and radiative luminosity $L$.   
The resulting luminosity provides the photoionisation heating rate inside the bubble, which we compare with the radiative cooling rate to iteratively solve the chemo-thermal equilibrium state.

We initially neglect the hyper-Eddington regime and compute the accretion rate and luminosity from either the BHL rate within the bubble, $\mdotBHLII(\mBH,\nH,T,v,\TII)$, or the Eddington rate $\mdotEdd$, defined by
\bea
\LEdd &=& \fr{4\pi G \mH c}{\sigmaT} \mBH \nn\\
\mdotEdd &=& \fr{1-\eta}{\eta}\fr{\LEdd}{c^2} \equiv \fr{\mBH}{\tEdd} \nn\\
&=& 2.0 \x 10^{-5} \Msuny ~\mBHsan  \\ 
\tEdd &=& \fr{c \sigmaT}{4\pi G \mH c}\fr{\eta}{1-\eta} = 5.0 \x 10 \Myr 
\eea
Here we adopt a radiative efficiency $\eta = 0.1$; $\sigmaT$ is the Thomson cross-section.

We then set
\bea
\mdotBH &=&
\begin{cases}
    \mdotEdd & \text{if \ $\mdotBHLII > \mdotEdd$,}\\
    \mdotBHLII & \text{otherwise}
\end{cases} \\[4pt]
L &=&
\begin{cases}
    \LEdd & \text{if \ $\mdotBHLII > \mdotEdd$,}\\
    \eta\,\mdotBHLII\,c^{2} & \text{otherwise.}
\end{cases}
\eea

Using the values of $\mBH$, $\nHII$, and $\TII$, the radius of the ionised bubble (Strömgren radius) is
\bea
\rStr &=& \kmb{\fr{3\krb{L/\kag{\hnu}}}{4\pi \aB(\TII) \neII^{2}}}^{1/3} \nn\\
&=& \kmb{ \fr{3\krb{L/\kag{\hnu}}}{4\pi \aB(\TII) (\rhoII / 2\muII)^{2}}}^{1/3}.
\eea
Note that the pressure equilibrium depends on the total number density of ions and electrons, $\nII=\rhoII/\muII$, whereas the recombination rate depends only on the number density of electrons (or of ionised hydrogen), $\neII\simeq\rhoII/2\muII$; hence the factor of~2 in the denominator.  
Assuming a spectral energy distribution $\Lv \propto \nu^{-1.5}$, we adopt a mean ionising-photon energy of $\kag{\hnu} = 3 \x 13.6 \eV$.  
The case-B recombination coefficient is
\bea
\aB(\TII) = 2.6 \x 10^{-13} \cmcs \norm{\TII}{10^4\K}{-0.8}. \label{eq:case-B}
\eea

This Strömgren radius is compared with the BHL radius calculated from the ambient neutral gas:
\bea
\rBHL = \fr{2 G \mBH}{\cs^{2} + v^{2}}.
\eea
Photon trapping, and hence hyper-Eddington accretion, becomes possible only when $\rBHL > \rStr$.

The photon-trapping criterion for $L=\LEdd$ is
\bea
\rBHL &>& \rStr \nn\\
\Rar \nH &>& 2.4 \x 10^{6} \cc \norm{T}{10^{4}\K}{1/2} \mBHsan^{-1} \sqrt{1 + \fr{v^{2}}{\cs^{2}}} \nn\\
&& \times \norm{\TII}{4 \x 10^{4}\K}{1.4}, \label{eq:hyper-Eddington_condition}
\eea
which can be rewritten by comparing the Eddington accretion rate $\mdotEdd$ with the BHL rate $\mdotBHL = {4 \pi G^{2} \mBH^{2} \rho}/{(\cs^{2} + v^{2})^{3/2}}$ as
\bea
\fr{\mdotBHL}{\mdotEdd} &>& 1.6 \x 10^{3} \nn\\
&\x& \krb{1 + \fr{v^{2}}{\cs^{2}}}^{-1}
 \norm{\TII}{4\x 10^{4}\K}{0.4} \normi{\TII/4\x 10^{4}\K}{T/10^{4}\K}.\nn\\
 &&\label{eq:hyper-Eddington_condition_mdot}\\
\LRar ~~\tEdd&>& 8.2\x10 \normi{\Mcl}{\mBH}\tff\nn\\
&\x&\krb{1 + \fr{v^{2}}{\cs^{2}}}^{1/2}
 \norm{\TII}{4\x 10^{4}\K}{0.4} \normi{\TII/4\x 10^{4}\K}{T/10^{4}\K}.\nn\\
 &&\label{eq:hyper-Eddington_condition_tff}
\eea

When this criterion is satisfied, we adopt the following mass accretion rate.  
\citet{Hu+2022} and \citet{Toyouchi+2024} argue that, in hyper-Eddington regime, most of the inflowing gas in a slim disc is expelled by outflows, implying an effective BH growth rate
\bea
\mdotBH \simeq \krb{\mdotEdd\,\mdotBHL}^{1/2}.
\eea
The corresponding radiative luminosity is
\bea
L \simeq \LEdd \ln\normi{\mdotBHL}{\mdotEdd}. \label{eq:L_radiation}
\eea

The full prescription is therefore
\bea
\mdotBH =
\begin{cases}
    \krb{\mdotEdd\,\mdotBHL}^{1/2} & \text{if \ $\rStr < \rBHL$,}\\[4pt]
    \mdotEdd                      & \text{else if \ $\mdotBHLII > \mdotEdd$,}\\[4pt]
    \mdotBHLII                    & \text{otherwise,}
\end{cases}  \label{eq:mdot_prescription}
\eea
\bea
L =
\begin{cases}
    \LEdd \ln\normi{\mdotBHL}{\mdotEdd} & \text{if \ $\rStr < \rBHL$,}\\[4pt]
    \LEdd                               & \text{else if \ $\mdotBHLII > \mdotEdd$,}\\[4pt]
    \eta\,\mdotBHLII\,c^{2}             & \text{otherwise.}
\end{cases}
\eea

With the gas properties and luminosity, we solve the chemical equilibrium of (HI, HII, HeI, HeII, HeIII, and e) including photoionisation and collisional recombination.  
Using the resulting composition we compute the heating and cooling rates.  
Heating processes include photoionisation by BH radiation, whereas cooling processes comprise Ly$\alpha$, He$\alpha$, collisional ionisation of H and He, radiative recombination, and free-free emission.  
The equilibrium temperature obtained, $\TII \simeq 4\ft 5 \times 10^{4}\K$ in most runs, is broadly consistent with the detailed radiation-hydrodynamic simulations of \citet{Sugimura&Ricotti+2020}.

When $\mdotBHLII > \mdotEdd$ in Equation~(\ref{eq:mdot_prescription}), the accretion rate can exceed the Eddington rate for a short period with $\mdotBH=\mdotBHLII > \mdotEdd$.  
However, such episodes are self-regulated by radiation feedback at $L\simeq\LEdd$, and the super-Eddington phase lasts only $\sim10^{2\ft3}\yr$ \citep{Milosavljevic+2009b, Inayoshi+2016}.  
Over longer timescales the effective accretion rate is therefore limited to $\mdotBH\simeq\mdotEdd$.

Although Equation~(\ref{eq:L_radiation}) yields a luminosity formally smaller than the mechanical luminosity $L \sim \LEdd \krb{\mdotBHL/\mdotEdd}^{1/2}$ suggested by \citet{Hu+2022}, we adopt it as a conservative yet physically consistent choice (see Section~\ref{subsec:mechanical}).  
Momentum-driven feedback from the outflow, which could disrupt the clump, is neglected here and will be discussed in Section~\ref{subsec:mechanical}.

\section{BH-CLUMP KINEMATICS}\label{Appx:Kinematics}

We model the BH-clump system as a two-body problem.  
The BH is treated as a point mass, whereas the clump is approximated as a uniform sphere; they interact through gravity and additional forces introduced below.  
Forces acting on the BH, such as dynamical friction and the acceleration produced by the ionised bubble, are modelled in accordance with the local ionised-gas properties and the radiation feedback (see Appendices~\ref{Appx:Park-Ricotti_model} and~\ref{Appx:BH_accretion_model}).

We work in the centre-of-mass frame of the BH-clump system and follow the kinematic evolution of their relative motion.  
Defining $\bmx = \bmx_{\rm BH} - \bmx_{\rm cl},~ \bmv=\dot{\bmx},~r=|\bmx|,~ v=|\bmv|$ the relative coordinate, relative velocity, and their absolute values, we solve
\be
\Mcl \ddot{\bmx}_{\rm cl} = +\frac{G\Mcl(<r)\mBH}{r^{3}}\bmx  
+ \theta(\Rcl - r)\,\bigl(\fDF + \facc - \fshell\bigr)\frac{\bmv}{v}
\ee
\be
\mBH \ddot{\bmx}_{\rm BH} = -\frac{G\Mcl(<r)\mBH}{r^{3}}\bmx  
- \theta(\Rcl - r)\,\bigl(\fDF + \facc - \fshell\bigr)\frac{\bmv}{v},
\ee
where $\fDF$ is the dynamical-friction force, $\facc$ is the back-reaction force from mass accretion, and $\fshell$ is the accelerating force exerted by the ionised bubble. 
Here $\theta(t)$ is the Heaviside step function. 
These additional forces act when the BH stays inside the gas clump $(r<\Rcl)$.

Combining the two equations gives
\bea
\ddot{\bmx} = -\frac{G\Mtot(<r)}{r^{3}}\bmx  
- \theta(\Rcl - r)\,\mure^{-1}\bigl(\fDF + \facc - \fshell\bigr)\frac{\bmv}{v}. \label{eq:EoM}
\eea
The remaining quantities are
\bea
\Mtot &\equiv& \Mcl + \mBH, \quad
\mure \equiv \bigl(\Mcl^{-1} + \mBH^{-1}\bigr)^{-1}.
\eea
The enclosed mass within the clump is
\bea
\Mcl(<r) &\equiv& 
\begin{cases}
    \Mcl \left(r/\Rcl\right)^{3} & (r \leq \Rcl),\\
    \Mcl & (r > \Rcl),
\end{cases}\\
\Mtot(<r) &\equiv& 
\begin{cases}
    \Mtot \left(r/\Rcl\right)^{3} & (r \leq \Rcl),\\
    \Mtot & (r > \Rcl).
\end{cases}
\eea

The back-reaction force from accretion is
\bea
\facc = |\mBH \dot{v}| = \mdotBH v.
\eea

The dynamical-friction force adopts the form of \citet{TanakaHaiman2009}:
\bea
\fDF = \frac{4\pi G^{2}\mBH^{2}\rhoII}{\vII^{2}}\,F(\vII/\csII).\label{eq:fric_II}
\eea
Here we use the ionised-gas values $\vII, \csII, \text{and } \rhoII$ because the dominant contribution originates from it inside the BHL radius when $\rBHL < \rStr$.  
If the photon-trapping condition $\rBHL > \rStr$ is fulfilled, we instead employ the ambient neutral-gas values $v$, $\cs$, and $\rho$:
\bea
\fDF = \frac{4\pi G^{2}\mBH^{2}\rho}{v^{2}}\,F(v/\cs).\label{eq:fric_I}
\eea
The Mach-number-dependent function is
\bea
F(\Mach)=
\begin{cases}
    0.5\ln\Lambda\left[{\rm erf}\!\left(\frac{\Mach}{\sqrt{2}}\right)-\sqrt{\frac{2}{\pi}}\Mach e^{-\Mach^{2}/2}\right] & (0\le\Mach\le0.8),\\[0.5em]
    1.5\ln\Lambda\left[{\rm erf}\!\left(\frac{\Mach}{\sqrt{2}}\right)-\sqrt{\frac{2}{\pi}}\Mach e^{-\Mach^{2}/2}\right] & (0.8<\Mach\le1.5),\\[0.5em]
    0.5\ln\!\left(\frac{\Mach+1}{\Mach-1}\right)-\ln\Lambda & (\Mach>1.5).
\end{cases}\nn
\eea
\bea
&&
\eea
We set the Coulomb logarithm to $\ln\Lambda = 3.1$.

The acceleration due to the ionised shell, $\fshell$, follows \citet{Toyouchi+2020} and \citet{Sugimura&Ricotti+2020}:
\bea
\fshell &=& \frac{G}{\rStr^{2}}
\int_0^{\pi}\sin{\theta}\rmd \theta\int_0^{2\pi}\rmd \phi \krb{ \rStr^{2}\Drshell \rhoshell\cos{\theta}}   \nn\\
&=& \pi G \mBH \rho \rStr\left(1 - \frac{v^{2} + \cs^{2}}{2\csII v}\right),
\eea
with
\bea
\rhoshell &\simeq& \left(\frac{v}{\cshell}\right)^{2}\rho,\\
\Drshell &\simeq& \rStr\left(\frac{\cshell}{v}\right)^{2}\!\left(1 - \frac{v^{2} + \cs^{2}}{2\csII v}\right).
\eea
Here $\cshell$ is the sound velocity in the shell.  
Following \citet{Sugimura&Ricotti+2020, Toyouchi+2020}, we set $\fshell = 0$ if any of the following conditions are satisfied:
\begin{itemize}
    \item the ionisation front is not D-type ($v<\cs$ or $\csII<v$);
    \item the shell thickness obeys $\Drshell < 0.05\rStr$ and $\TII / T > 3$;
    \item the photon-trapping condition $\rBHL > \rStr$ holds.
\end{itemize}

We integrate the relative motion in two dimensions.  
The BH is initially unbound from the clump; its conditions at infinity are parametrised by $(\vin, b)$.  
We solve the Keplerian trajectory analytically from infinity to obtain the initial state at first entry into the clump surface, $r=\Rcl$, as
\bea
x_{r}(t_{0}) &=& \Rcl,\quad x_{\theta}(t_{0}) = 0,\\
v_{r}(t_{0}) &=& \sqrt{\vin^{2}\!\left(1 - \frac{b^{2}}{\Rcl^{2}}\right)+\fr{2G\Mtot}{\Rcl}},\\
v_{\theta}(t_{0}) &=& \frac{b}{\Rcl}\,\vin.
\eea

We additionally compute the specific mechanical energy, defined per unit reduced mass in the centre-of-mass frame, as
\bea
E &\equiv& \fr{\bmv^{2}}{2} - \fr{G\Mcl(<r)\mBH}{\mure r} =  \fr{v^{2}}{2} - \fr{G\Mtot(<r)}{r},
\eea
and use it to evaluate whether the BH is gravitationally bound to the clump ($E<0$) or remains unbound ($E>0$).

\end{document}